\documentclass[10pt, aps, prd, twocolumn, superscriptaddress, showpacs]{revtex4-1}

\usepackage{graphicx}
\usepackage{amsmath, amssymb}

\usepackage{hyperref}
\usepackage{setspace, booktabs, tabularx, multirow}

\hypersetup{
    pdfnewwindow=true,  
    colorlinks=true,          
    linkcolor=blue,           
    citecolor=blue,           
    filecolor=blue,            
    urlcolor=blue              
}

\begin{document}

\title{Demystifying the PeV Cascades in IceCube: Less (Energy) is More (Events)} 

\author{Ranjan Laha}
\affiliation{Center for Cosmology and AstroParticle Physics (CCAPP), Ohio State University, Columbus, OH 43210}
\affiliation{Department of Physics, Ohio State University, Columbus, OH 43210}

\author{John F. Beacom}
\affiliation{Center for Cosmology and AstroParticle Physics (CCAPP), Ohio State University, Columbus, OH 43210}
\affiliation{Department of Physics, Ohio State University, Columbus, OH 43210}
\affiliation{Department of Astronomy, Ohio State University, Columbus, OH 43210}

\author{Basudeb Dasgupta}
\affiliation{International Center for Theoretical Physics, 34014 Trieste, Italy}

\author{Shunsaku Horiuchi}
\affiliation{Center for Cosmology, Department of Physics and Astronomy, University of California, Irvine, CA 92697}

\author{Kohta Murase}
\affiliation{Hubble Fellow, School of Natural Sciences, Institute for Advanced Study, Princeton, NJ 08540 \\ 
{\tt laha.1@osu.edu, beacom.7@osu.edu, bdasgupta@ictp.it, s.horiuchi@uci.edu, murase@ias.edu} \smallskip}

\date{July 29, 2013}

\begin{abstract}
The IceCube neutrino observatory has detected two cascade events with energies near 1 PeV~\cite{Ishihara:2012nu, Aartsen:2013bka}.  Without invoking new physics, we analyze the source of these neutrinos.  We show that atmospheric conventional neutrinos and cosmogenic neutrinos (those produced in the propagation of ultra-high-energy cosmic rays) are strongly disfavored.  For atmospheric prompt neutrinos or a diffuse background of neutrinos produced in astrophysical objects, the situation is less clear.  We show that there are tensions with observed data, but that the details depend on the least-known aspects of the IceCube analysis.  Very likely, prompt neutrinos are disfavored and astrophysical neutrinos are plausible.  We demonstrate that the fastest way to reveal the origin of the observed PeV neutrinos is to search for neutrino cascades in the range below 1 PeV, for which dedicated analyses with high sensitivity have yet to appear, and where many more events could be found.
\end{abstract}



\maketitle


\section{Introduction}

Neutrino astronomy has long promised to reveal the astrophysical sites of particle acceleration and the nature of cosmic rays~\cite{Greisen:1960wc, Pontecorvo:1963wk, Lande:1979gr, Gaisser:1994yf, Learned:2000sw, Halzen:2002pg, Lipari:2006uw, Becker:2007sv, Anchordoqui:2009nf}.  The lack of adequately-sized neutrino detectors has been a deterrent in turning this dream into reality. The recent completion of the IceCube detector has raised hope of addressing these long-standing problems~\cite{IceCube:2011ab}.  Encouraging this hope, an analysis of very high energy neutrino events in the IceCube detector during 2010--2012, as construction was finishing, found two candidate neutrino cascade events with energies near 1 PeV~\cite{Ishihara:2012nu, Aartsen:2013bka}.

These are the highest energy neutrinos ever detected -- they are $10^6$ times more energetic than typical GeV atmospheric neutrinos.  They signal the entry of neutrino astronomy into the PeV era, made possible by the huge size of IceCube.  However, these events have led to several mysteries.  Where did they come from?  Although we expect $\nu_\mu$ to be more detectable than $\nu_e$ due to the long range of the muons, why are there two cascade events and zero muon track events?  Why are the two event energies so close to each other and to the analysis threshold?  Is the neutrino flux required to explain these events consistent with previous limits and with other data?

These PeV neutrino events have spurred a flurry of activity, due to the importance of the potential first discovery of non-atmospheric high-energy neutrinos.  Astrophysical neutrinos -- those produced inside distant sources -- have been considered~\cite{Cholis:2012kq, Liu:2012pf, Kistler:2013my, He:2013cqa, Kalashev:2013vba, Gupta:2013xfa, Fox:2013oza, Stecker:2013fxa, Anchordoqui:2013lna, Murase:2013rfa, Gao:2013rxa, Murase:2013ffa}.  Cosmogenic neutrinos -- those produced in the propagation of ultra-high-energy cosmic rays -- have also been considered~\cite{Roulet:2012rv, Fargion:2012mw, Fargion:2012zx, Fargion:2012zc}.  Other papers have proposed more exotic explanations~\cite{Pakvasa:2012db, Baerwald:2012kc, Bhattacharya:2012fh, Feldstein:2013kka, Barger:2013pla}.  Novel tests of the data or of new physics have been noted~\cite{Borriello:2013ala, Vissani:2013iga, Stecker:2013jfa}.

We provide a new general analysis of the source of these two events,  focusing on the simplest and most straightforward scenarios, and including many realistic aspects of neutrino detection in IceCube (for our early results, see Refs.~\cite{LahaAspen:2013, LahaIPA:2013}).  We assume that both events were neutrino-induced and that neutrinos have only standard properties and interactions.  We assess which scenarios are compatible with the present data and the implications of this discovery.  Importantly, we detail how these scenarios can be tested by new analyses.

The flux of atmospheric conventional neutrinos at PeV energies is much too low to give rise to these two cascade events.  Cosmogenic neutrinos are also very unlikely to be the source, due to the lack of higher-energy events.  Atmospheric prompt neutrinos do not appear to be a plausible source, but they should not be dismissed lightly.  A diffuse background of neutrinos from astrophysical objects can reasonably explain the observed data, though there are strong constraints on the spectrum.  A full assessment of these models will require more details about the IceCube search strategies.

New analyses optimized for energies near and below 1~PeV are urgently needed.  The cascade or shower channel for electron neutrinos is especially important, because its atmospheric conventional neutrino background is much lower than for muon neutrinos, as first shown by Beacom and Candia in 2004~\cite{Beacom:2004jb}.  There are great opportunities to better exploit this detection channel.

In Sec.~II, we begin with the basic information on these two PeV cascade events and what it suggests, which we support with quantitative details in later sections.  In Sec.~III, we test whether various neutrino fluxes can be the source of these two events.  In Sec.~IV, we detail how searches for cascades and tracks in the energy range below 1 PeV will robustly distinguish between various sources.  We conclude in Sec.~V, including commenting on preliminary new IceCube events below 1 PeV.


\section{What is known about the events}

These two events were detected as PeV cascades during the 2010--2012 runs.  They were identified in the extremely high energy (EHE) search, which is optimized for the detection of EeV $= 10^3$ PeV cosmogenic neutrinos~\cite{Aartsen:2013bka}.  This search has strong cuts to decisively reject detector backgrounds, and these cuts greatly affect the acceptance for signal events, especially in the PeV range, which is the edge of the considered energy range, because relatively few cosmogenic events are expected there.

Our analysis focuses on the PeV range and below.  This section introduces the events and their implications.  The reconstructed event energies are $1.04 \pm 0.16$ PeV and $1.14 \pm 0.17$ PeV~\cite{Aartsen:2013bka}.  This disfavors neutrino interactions at the Glashow resonance at 6.3 PeV, for which the cascade energy should generally be the same; we discuss exceptions below.  The absence of higher-energy events disfavors cosmogenic neutrinos, as their detection probability is largest in the EeV range. 

The values of the energies, and especially their proximity to each other, are crucial.  We assume that the detected energies are probable values in the distribution of possible values; this is reinforced by there being two similar events.  The minimal explanation of the two energies is that this distribution is peaked at $\sim 1$ PeV, due to a drop in detector acceptance at lower energies and decreasing neutrino spectra at higher energies.  The analysis threshold for this search is $\sim 1$ PeV~\cite{Aartsen:2013bka}, which makes it remarkable that both events were detected there.  Very likely, there are already many additional signal events to be found at lower energies, but isolating them will require new searches with cuts optimized for cascades in the PeV range.  Events will likely also be found at higher energies, but this will take additional exposure time.

The types of events -- two cascades, zero muon tracks, and zero tau-lepton events -- also arise from the nature of the search criteria, which are primarily based on the total number of detected photoelectrons.  In addition, downgoing track-like events are strongly suppressed by the cuts.  The effective area curves for different flavors show that this search strategy gives the maximum exposure in the energy range 1--10 PeV to $\nu_e + \bar{\nu}_e$~\cite{Aartsen:2013bka}.   The efficiency for $\nu_\mu + \bar{\nu}_\mu$, which should be more detectable due to the long range of the muons, is suppressed, because the muons do not deposit their full energy in the detector.  The efficiency for $\nu_\tau + \bar{\nu}_\tau$ is suppressed because of the tau-lepton decay energy carried by neutrinos.  This explains the non-observation of muon track and tau-lepton events; future searches can be optimized to find them.
 
The most likely scenario is that both cascade events arise from charged current (CC) interactions of $\nu_e + \bar{\nu}_e$, for which the detectable cascade energy is nearly the full neutrino energy.  Because of the above suppressions, we neglect the rare cases in which $\nu_\mu + \bar{\nu}_\mu$ or $\nu_\tau + \bar{\nu}_\tau$ CC events resemble $\nu_e + \bar{\nu}_e$ cascades, due to the muon getting a small fraction of the neutrino energy or the tau lepton decaying quickly.  Neutral current (NC) interactions of all flavors of neutrinos also give cascades.  The cross section is 2.4 times smaller near 1 PeV, though three neutrino flavors may contribute.  The more important point is that the average cascade energy in a NC interaction is only $\sim 0.25$ of the neutrino energy in the PeV range, which makes the event much less detectable~\cite{Aartsen:2013bka}.  It is unlikely that NC interactions could be the source of these events, especially both of them, because the cascade energies are so close to each other and the analysis threshold.

These events are consistent with a steady, isotropic diffuse source, and we assume this, though other possibilities are not excluded.  The events were separated temporally by 5 months; the search ran for about 2 years.  It is difficult to measure the directions of cascade events, as the signal regions in the detector are large and spherelike.  No event directions are reported in the IceCube paper~\cite{Aartsen:2013bka}, and preliminary IceCube results from conferences vary significantly~\cite{Whitehorn:tevpa2012, Neilson:Pheno2013}.  Future analyses are expected to have an angular resolution of $\sim 10$ degrees for cascades near 1 PeV (and worse at lower energies)~\cite{Whitehorn:tevpa2012}.  For upgoing events that pass through Earth's core, with a zenith angle greater than $\sim 150^\circ$ ($\sim 7\%$ of the full sky), there would be especially significant attenuation due to interactions in Earth~\cite{Abbasi:2010ak, Abbasi:2011ji}.  Prompt neutrinos that are sufficiently downgoing will be accompanied by cascades that trigger the IceTop surface detector~\cite{Ishihara:2012nu,Schonert:2008is}; this was not seen, and studies of its efficiency are ongoing.

\begin{figure}[t]
\includegraphics[width=\columnwidth]{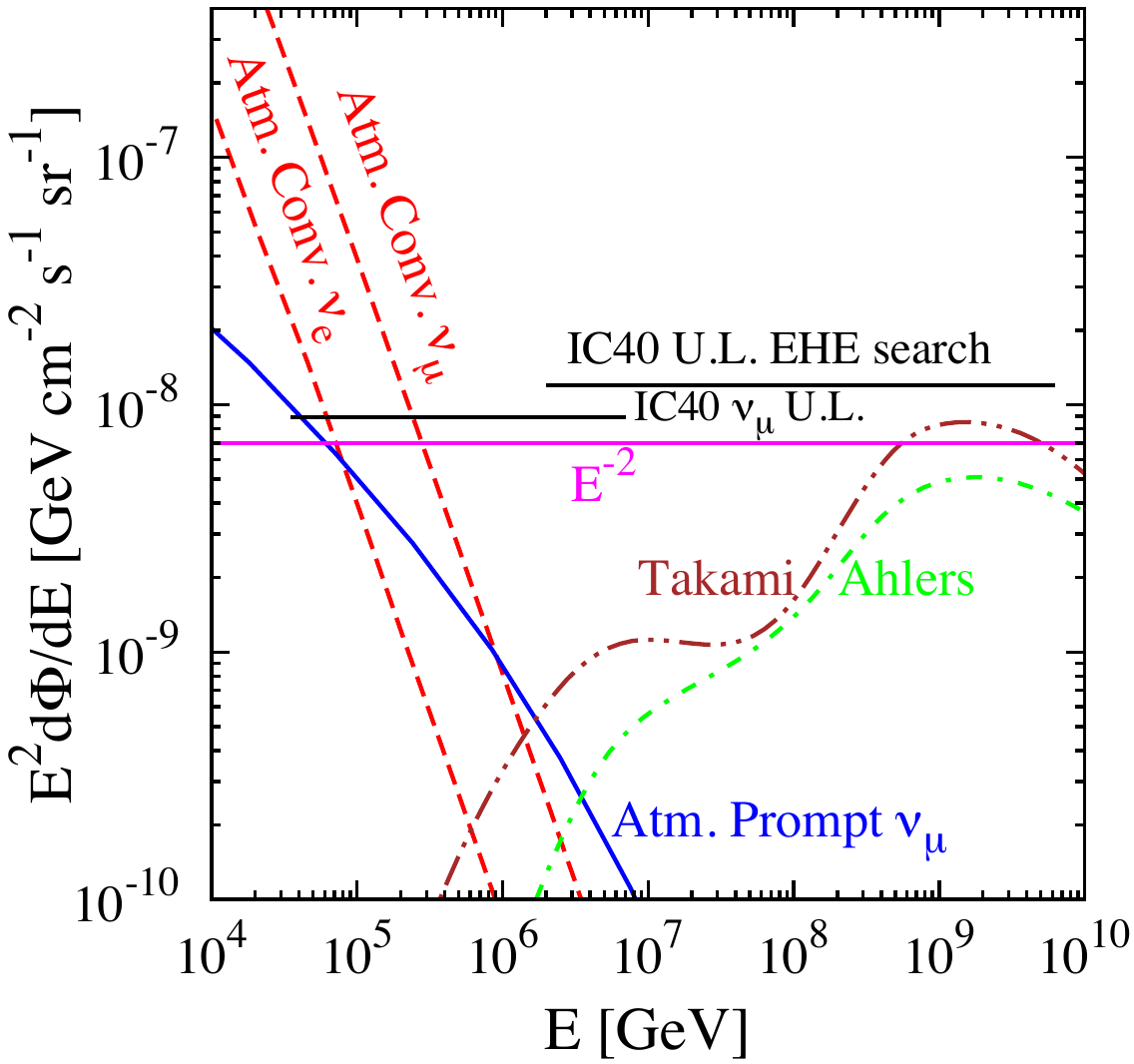}
\caption{Neutrino fluxes as a function of neutrino energy. The atmospheric conventional $\nu_{\mu} + \bar{\nu}_{\mu}$ and $\nu_e + \bar{\nu}_e$ spectra are from Ref.~\cite{Honda:2006qj, Abbasi:2011jx}.  The atmospheric prompt $\nu_{\mu} + \bar{\nu}_{\mu}$ spectrum (the $\nu_{e} + \bar{\nu}_{e}$ flux is the same) is the Enberg (std.)\ model~\cite{Enberg:2008te}.  Example cosmogenic EHE neutrino fluxes ($\nu + \bar{\nu}$ for one flavor) are from Refs.~\cite{Takami:2007pp, Ahlers:2010fw}.  An $E^{-2}$ astrophysical neutrino spectrum for one flavor of $\nu + \bar{\nu}$, normalized as discussed below, is shown, along with current upper limits from IceCube~\cite{Abbasi:2011jx, Abbasi:2011ji}.}
\label{fig:FluxOverview}
\end{figure}

Figure~\ref{fig:FluxOverview} shows some relevant neutrino spectra.

\section{What can be the source?}

In this section, we first discuss our general approach to testing possible spectra, given that much is not yet known.  We then discuss cascade detection in IceCube, followed by detailed discussions of possible sources of these events and a summary of remaining issues.


\subsection{Our approach to assessing source spectra}

The two PeV events were found in the EHE search, which is not optimized for detection in the PeV energy range.  The cuts required to reject backgrounds reduce the probability of detecting signal events, especially at these relatively low energies.  The effective area plot in Ref.~\cite{Aartsen:2013bka} shows that the neutrino detection probability falls very quickly with decreasing neutrino energy, plummeting below $\sim 1$ PeV.  In the range 1--10 PeV, the variation of this probability with energy is far too rapid to be accounted for by the variation of the neutrino cross section.  The difference is due to strong event selection cuts.

We first follow a ``theorist's approach" to calculating the event rates, using the flux, cross section, Earth attenuation, and other factors.  We are unable to reproduce the effective area for the $\nu_e + \bar{\nu}_e$ flavors~\cite{Aartsen:2013bka}.  A straightforward calculation -- not including the effects of the strong cuts -- is about one order of magnitude larger than the effective area of Ref.~\cite{Aartsen:2013bka} near 1 PeV, and this point has not been noted before.  (We can reproduce the effective area for other IceCube searches, e.g., Ref.~\cite{Abbasi:2011jx}.)  However, as both events were detected at $\sim 1$ PeV, there should be an appreciable detection probability there.

In the following, we show event spectra calculated using this ``theorist's approach" as well as with the effective area from Ref.~\cite{Aartsen:2013bka}.  Our results are adequate to make preliminary assessments of which sources could give rise to these events, though the hypothesis likelihoods are uncertain.  Further, we have enough information to make predictions for how to test the origin of these events.  Given the large uncertainties on the inputs, we make various approximations at the level of a few tens of percent.

Figure~\ref{fig:FluxAstro} shows the main spectra we consider for explaining the PeV events (details are given below).  The measured atmospheric conventional neutrino data should be taken with some caution.   Assumptions were made to work backwards from detected energy to neutrino energy, especially for the muon tracks, and the error bars are highly correlated.  In addition, the publication of detected cascade events is relatively new, and measured atmospheric neutrino cascade spectra reach only as high as 10 TeV~\cite{Aartsen:2012uu}.  In between there and 1 PeV lies an important opportunity for discovery in a short time, likely by improved analyses of existing data.

A first tension appears in the normalization of a possible source spectrum.  If it is too large, then this would conflict with measurements of atmospheric neutrino data, which largely agree with predictions.  If it is too small, then this would conflict with the observation of the two PeV events.  We choose acceptable normalizations in Fig.~\ref{fig:FluxAstro} and later estimate the probabilities of detecting two events in the PeV range.  The normalizations could be increased, given the large uncertainties; the power-law fluxes could be increased by about a factor of 2, and the prompt flux by more.  A second tension appears in the slope of a possible source spectrum.  If it is too steep, then the spectrum will exceed measurements of atmospheric conventional neutrinos at lower energies unless the spectrum breaks.  If it is not steep enough, then it will have too many events expected above 1 PeV.

For both of these issues, the degree of statistical tension would be calculable in a full analysis, whereas here we can only estimate it.  We consider two energy bins; these were chosen post hoc, but the fact the event energies are so close to each other and the threshold at 1 PeV seems to be a strong clue.  The first bin is 1--2 PeV, which easily contains both points within energy uncertainties.  Detections at lower energies are assumed impossible due to the threshold.  Detections at higher energies are considered with a second bin, 2--10 PeV; for falling spectra, the exact value of the upper limit is not very important.

We present our results in terms of detectable energy, which is not always the same as neutrino energy, as explained below.  This is closer to what is actually measured, allowing for much better control in separating signals and backgrounds.

\begin{figure}[t]
\includegraphics[width=\columnwidth]{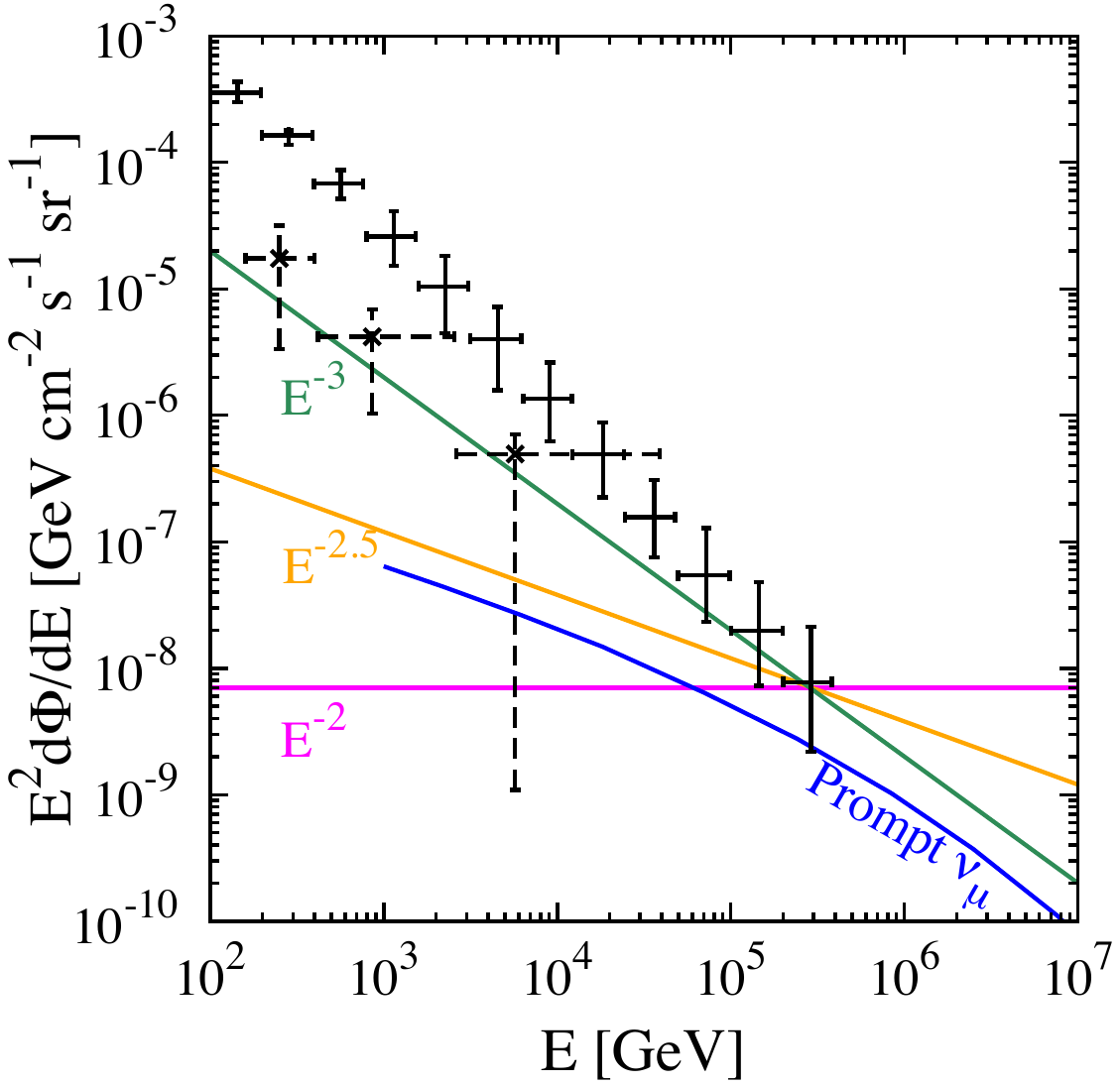}
\caption{Example neutrino fluxes (for one flavor of $\nu + \bar{\nu}$) that might produce the PeV events, compared to the atmospheric conventional $\nu_{\mu} + \bar{\nu}_{\mu}$ (upper points) and $\nu_{e} + \bar{\nu}_{e}$ (lower points) fluxes measured by IceCube~\cite{Abbasi:2010ie, Aartsen:2012uu}.  The power-law astrophysical fluxes are normalized so that they do not exceed the measured data.  The atmospheric prompt neutrino flux is only shown above 1 TeV, following Ref.~\cite{Enberg:2008te}.}
\label{fig:FluxAstro}
\end{figure}


\subsection{Cascade detection in IceCube}

The neutrino-nucleon cross sections $\sigma(E_\nu)$ near 1 PeV are well known~\cite{Gandhi:1998ri, Connolly:2011vc, CooperSarkar:2011pa, Block:2013nia}.  In CC cascade events initiated by $\nu_e + \bar{\nu}_e$, the neutrino interacts with a nucleon, leading to a hadronic cascade, and produces an electron or positron, leading to an electromagnetic cascade.  The division of the neutrino energy $E_\nu$ depends on the inelasticity $y$, for which $\langle y \rangle \simeq 0.25$ near 1 PeV and varying slightly with energy~\cite{Gandhi:1995tf}.  The outgoing lepton has energy $(1 - \langle y \rangle) E_\nu$, with the remainder going to the hadrons, so that the detectable total cascade energy is $\simeq E_\nu$.  The cascade leads to a roughly spherical distribution of hit phototubes over a diameter of a few$\times 100$ m, though the cascade size is several meters.  Cascades produced by the NC interactions of all flavors are similar, though the hadronic cascade energy is just $\langle y \rangle E_\nu$ instead of $E_\nu$, so NC cascades can normally be neglected for all but atmospheric conventional neutrinos~\cite{Beacom:2004jb}.

In the ``theorist's approach" or ideal case, the event rate spectrum for $\nu_e + \bar{\nu}_e$ CC cascades is
\begin{eqnarray}
\frac{dN}{dE_{casc}}
& & \simeq 2\pi \, \rho \, N_A \, V \, T \\
& \times & \int^{+1}_{-1} d(\cos{\theta_z}) \, \dfrac{d\Phi}{dE_\nu}(E_\nu) \, \sigma(E_\nu) \, e^{-\tau(E_\nu, \cos{\theta_z})} \,. \nonumber
\label{eq:chargedcurrentcascade}
\end{eqnarray}
The number of target nucleons is $\rho \,N_A \,V$, where $\rho$ is the ice density (in g cm$^{-3})$, $N_A$ the Avogadro number, and the IceCube volume is $V \simeq 1$ km$^3$.  The observation time is $T = 615.9$ days~\cite{Aartsen:2013bka}.  The neutrino cross section $\sigma$ (in cm$^2$) and the neutrino intensity spectrum $d\Phi/dE_\nu$ (in GeV$^{-1}$ cm$^{-2}$ s$^{-1}$ sr$^{-1}$) are evaluated at $E_\nu \simeq E_{casc}$ (in GeV).  Neutrino flux attenuation en route to the detector, which depends on energy and  zenith angle, is taken into account in the optical depth $\tau = \ell/\lambda$ assuming a constant density of 3 g cm$^{-3}$ for Earth, where $\ell$ is the path length and $\lambda$ the mean free path.  We include NC interactions via simple modifications to the above, including a factor $1/\langle y \rangle$ due to the change in the energy differential.

The CC cross section varies smoothly with energy, except near the Glashow resonance at 6.3 PeV, which is caused by the resonant production of an on-shell $W$ boson by $\bar{\nu}_e + e^- \rightarrow W^-$~\cite{Glashow:1960zz, Gandhi:1998ri}.  The $W$ decays promptly, typically depositing most of its energy in the detector.  About 10\% of the time, the decay to an electron and an antineutrino leads to a range of smaller deposited energies; assuming that there are enough such interactions, the probability for this to happen twice is thus $\lesssim 1\%$~\cite{Barger:2012mz}.  At 6.3 PeV, the ratio of the cross section for $\bar{\nu}_e$ to interact with an electron instead of a nucleon is 350~\cite{Gandhi:1998ri, Bhattacharya:2012fh}.  The overall importance of this is reduced by an equal flux of $\nu_e$, half as many electron as nucleon targets, and the opacity of Earth to $\bar{\nu}_e$ at this energy.  In the effective area plot of Ref.~\cite{Aartsen:2013bka}, the enhancement is thus only a factor of $\simeq 15$ in a bin of width $\Delta(\log E) = 0.05$.

The CC cascade events initiated by $\nu_\tau + \bar{\nu}_\tau$ can be similar those those initiated by $\nu_e + \bar{\nu}_e$.  At $\sim$ 1 PeV, the tau-lepton decay length is $\sim 50$ m.  (Above $\sim 5$ PeV, where the tau lepton travels far enough that the cascades from production and decay separate significantly, there are very distinct signatures~\cite{Learned:1994wg, Beacom:2003nh}.)  In tau-lepton decays, the fraction of energy lost to neutrinos is $\simeq 0.3$; the fraction of $E_\nu$ deposited for $\nu_\tau + \bar{\nu}_\tau$ events with prompt tau-lepton decays is then $\simeq \langle y \rangle + 0.7 (1 - \langle y \rangle) \simeq 0.8$ at PeV energies~\cite{Ritz:1987mh, Halzen:1998be}.  We do not include $\nu_\tau + \bar{\nu}_\tau$ events in our calculations of cascade spectra above 1 PeV for comparison with present data, but we do in our calculations below of possible future spectra below 1 PeV, which increases the rates by somewhat less than a factor of 2.

As a more realistic estimate, we calculate the cascade spectra using the effective area from Ref.~\cite{Aartsen:2013bka}, which leads to significantly smaller yields, due to the effects of the strong cuts in this search.  In this approach, the event rate spectrum for $\nu_e + \bar{\nu}_e$ cascades is
\begin{equation}
\frac{dN}{dE_{casc}} = 
4\pi \, A_{\rm eff} \, T \times \dfrac{d\Phi}{dE_\nu}(E_\nu) \, 
\label{eq:effectivearea}
\end{equation}
where $A_{\rm eff}$ takes into account all of the factors in Eqn.~(1) plus the detailed search cuts.

In both approaches, the effect of detector energy resolution on the spectrum must be taken into account.  We smooth the calculated spectra with a Gaussian of width $\delta E/E = 15\%$, taken to match the uncertainty on the energy of the two events.  Future analyses will likely have better energy resolution, more like 10\%~\cite{Whitehorn:tevpa2012}.  The effect of energy resolution on the Glashow resonance is especially significant, reducing its height and increasing its width while preserving the number of events.

\begin{figure*}[t]
\centering
\includegraphics[width=\columnwidth]{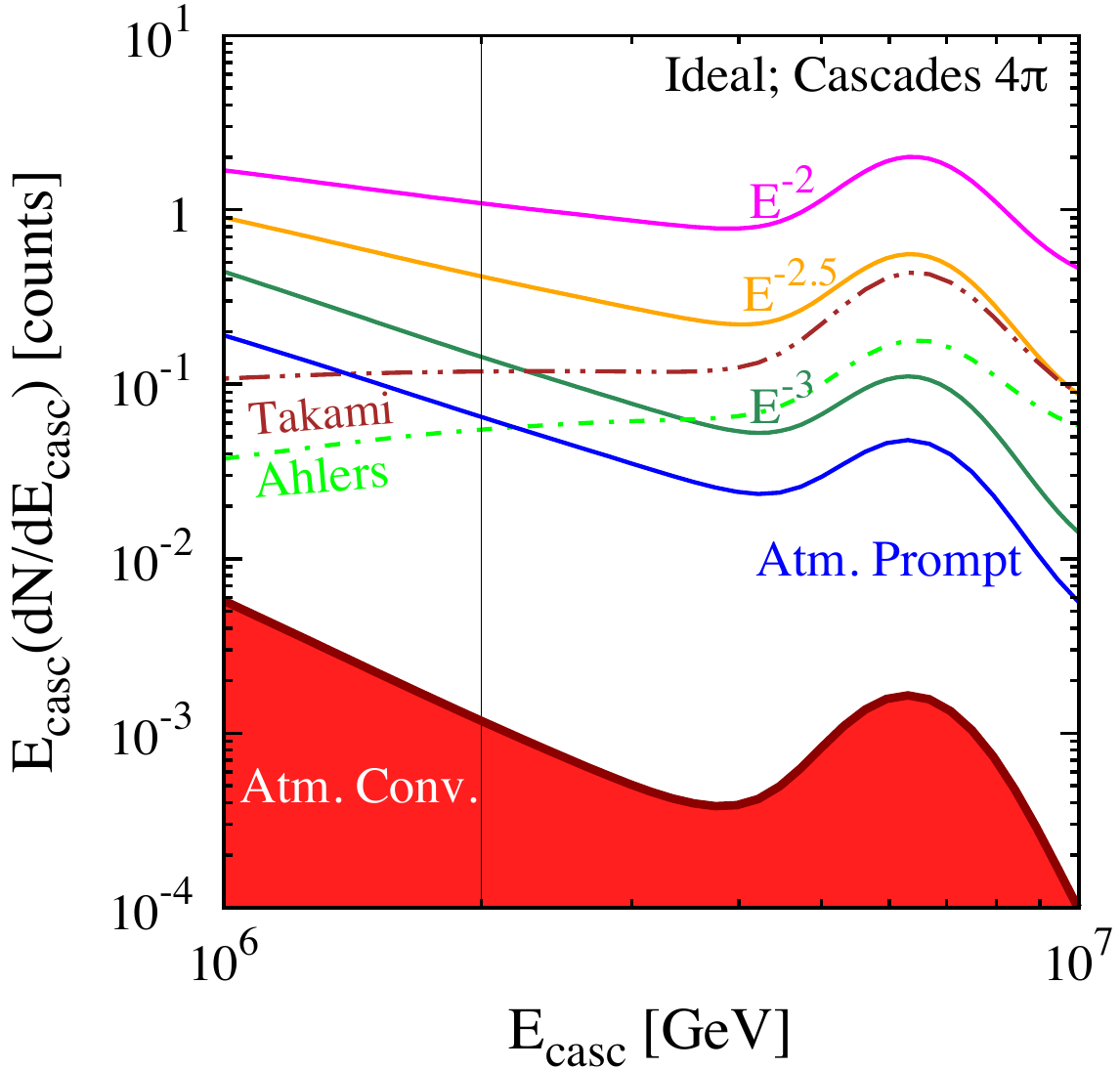}
\hspace{0.25cm}
\includegraphics[width=\columnwidth]{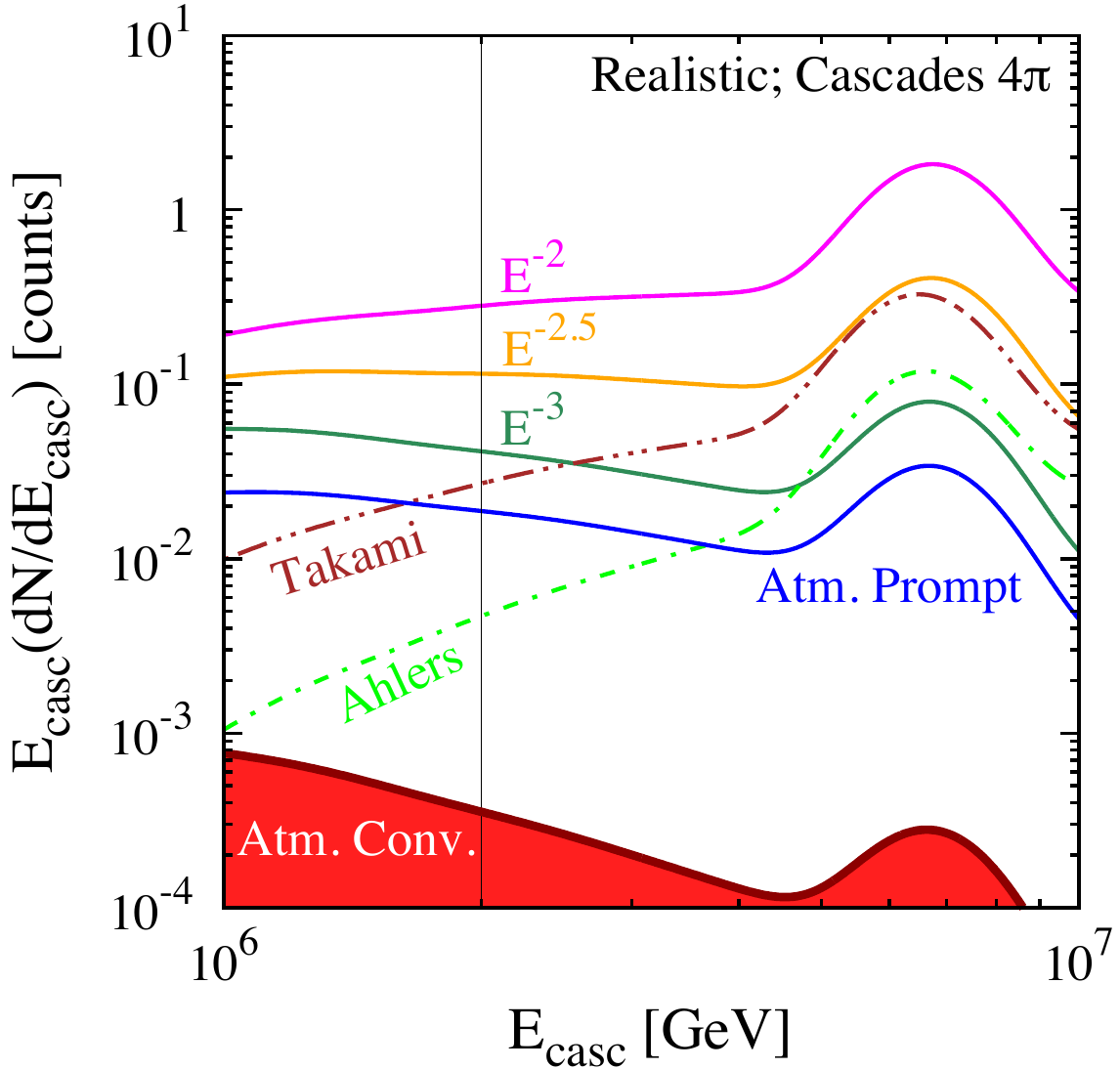}
\caption{$E dN/dE$ for neutrino-induced cascade spectra.  The {\bf left} panel is for the ideal case or ``theorist's approach," and the {\bf right} is for the realistic case using the effective area from Ref.~\cite{Aartsen:2013bka}.  These results are for the 615.9 days of exposure that included the two PeV events.  The power-law fluxes are normalized in Fig.~\ref{fig:FluxAstro}.  The thin vertical line denotes the boundary between our two bins.  The y-axis has a large logarithmic range to show several spectra.  The number of events in a region is proportional to the integrated area, i.e., to the height times the logarithmic energy range, so curves with low heights have very few events.}
\label{fig:Cascades}
\end{figure*}

Figure~\ref{fig:Cascades} shows our results (ideal and realistic) for the signal and background spectra.  The numbers of events in each bin for the realistic approach are given in Table~\ref{tab:presentyields}.

Energies in IceCube are measured with fractional, not fixed, precision, so $\log E$ is a more natural variable than $E$.  The number of bins of fixed width $dE = 1$ GeV in each decade of $\log E$ increases $\propto E$, so measured event spectra should then be presented as $E dN/dE = dN/d(\ln E) = 2.3^{-1} \, dN/d(\log E)$ instead of $dN/dE$.  Using $E dN/dE$ gives a correct visual representation of the relative detection probabilities in different ranges of $\log E$.  Further, this makes it much easier to estimate the area, i.e., the total number of events.  Using $E dN/dE$ and $\log E$ to estimate area means that both the height and width are dimensionless.  To get 1 event, the height must be $\sim 1$ over a moderate width.  For example, to estimate the number of events in the 1--2 PeV bin, multiply the height (averaged on an imagined linear y-axis) of a given curve by $d(\ln E) = 2.3 \, d(\log E) = \ln 2 = 0.69$.

In the remainder of this section, we first briefly state why it is unlikely that either atmospheric conventional neutrinos or cosmogenic neutrinos can explain the observed events.  We then provide more details on the results in Fig.~\ref{fig:Cascades}, focusing on the more promising scenarios, concluding with a discussion of the outstanding issues.

\begin{table}[b]
\caption{Expected numbers of cascade events in the two energy bins, obtained by integrating the curves in the right panel (the realistic approach using the effective area) of Fig.~\ref{fig:Cascades}.  These numbers are typically a factor of $\sim 5$ below those for the left panel (the ideal case or ``theorist's approach").}
\setlength{\extrarowheight}{4pt} 
\begin{ruledtabular}
\begin{spacing}{1.1}
\begin{tabular}{lcc}
Possible Source & ${\rm N}(1-2 {\rm\ PeV})$ & ${\rm N}(2-10 {\rm\ PeV})$ \\ 
\hline
Atm. Conv.~\cite{Honda:2006qj, Abbasi:2011jx} & 0.0004 & 0.0003 \\
\hline
Cosmogenic--Takami~\cite{Takami:2007pp} & 0.01 & 0.2 \\
Cosmogenic--Ahlers~\cite{Ahlers:2010fw} & 0.002 & 0.06 \\
\hline
Atm. Prompt~\cite{Enberg:2008te} & 0.02 & 0.03 \\
\hline
Astrophysical $E^{-2}$ & 0.2 & 1 \\
Astrophysical $E^{-2.5}$ & 0.08 & 0.3 \\
Astrophysical $E^{-3}$ & 0.03 & 0.06 \\
\end{tabular}
\end{spacing}
\end{ruledtabular}
\label{tab:presentyields}
\end{table}


\subsection{Atmospheric conventional fluxes: very unlikely}

Because atmospheric conventional neutrinos definitely exist, it is important to ask if they could produce these events.  We show the $\nu_\mu + \bar{\nu}_\mu$ and $\nu_e + \bar{\nu}_e$ fluxes from Ref.~\cite{Honda:2006qj, Abbasi:2011jx} in Fig.~\ref{fig:FluxOverview}.  The $\nu_\tau + \bar{\nu}_\tau$ flux is much smaller, because both direct production and neutrino oscillations at these energies are suppressed, and it is not shown.

In the muon track channel, the atmospheric conventional $\nu_\mu + \bar{\nu}_\mu$ flux is a significant background to new signals even at high energies.  However, as shown in Ref.~\cite{Beacom:2004jb}, the atmospheric conventional backgrounds for $\nu_e + \bar{\nu}_e$ are significantly less, which means that new signals can emerge at lower energies.  To see this, it is necessary to plot predicted event spectra in terms of detectable cascade energy instead of neutrino energy.  For $\nu_e + \bar{\nu}_e$ CC events, these are the same.  For NC $\nu_\mu + \bar{\nu}_\mu$ events, which have a small energy deposition, it is a big difference.  Going from Fig.~\ref{fig:FluxOverview} to the left panel of Fig.~\ref{fig:Cascades}, the importance of atmospheric conventional neutrinos relative to other sources (e.g., the $E^{-2}$ spectrum) is greatly reduced.  This is what makes cascade searches so powerful~\cite{Beacom:2004jb}.

The complete (CC + NC) $\nu_e + \bar{\nu}_e$ cascade spectrum from atmospheric conventional neutrinos is shown in Fig.~\ref{fig:Cascades}, with the integrated numbers of events for the realistic case given in Table~\ref{tab:presentyields}.  If we also include muon tracks (see below), the total number of events above 1 PeV increases to 0.008, which is consistent within uncertainties with the 0.012 of Ref.~\cite{Aartsen:2013bka}.  As these expected numbers are negligible, it is very unlikely that they can yield the PeV events.

Most downgoing atmospheric muons are easily identified as such.  In some rare cases, including muon bundles, these initiate events that look like neutrino-induced cascades.  The expected number of such events is 0.04~\cite{Aartsen:2013bka}, larger than the background from neutrinos.  All together, these conventional backgrounds have a $\sim 10^{-3}$ probability of producing at least two observed events.  These backgrounds can be studied further at lower energies, where they are larger.


\subsection{Cosmogenic neutrinos: very unlikely}

Cosmogenic neutrinos~\cite{Greisen:1966jv, Zatsepin:1966jv, Berezinsky:1970xj, Beresinsky:1969qj, Yoshida:1993pt, Engel:2001hd, Allard:2006mv, Anchordoqui:2007fi, Kotera:2010yn, Murase:2010gj} have been invoked as the source of the PeV events, in part because the EHE search was designed to detect them, albeit at much higher energies.  Example spectra~\cite{Takami:2007pp, Ahlers:2010fw} are shown in Fig.~\ref{fig:FluxOverview}.

The $\nu_e + \bar{\nu}_e$ cascade spectra are shown in Fig.~\ref{fig:Cascades} and the numbers of events are given in Table~\ref{tab:presentyields}.  Two problems are obvious.  First, the expected numbers of events are very small because the spectrum normalization is low.  Second, the predicted distribution of events emphasizes high, not low, energies.

The probability of having two or more $\nu_e + \bar{\nu}_e$ cascade events detected in the first bin is $\sim 10^{-4}$ for the model of Ref.~\cite{Takami:2007pp} and $\sim 10^{-6}$ for the model of Ref.~\cite{Ahlers:2010fw}.  There should also be a penalty factor to not have events in the second bin, but this is modest because the expected numbers of events are small.  For these models, there are comparable numbers of muon track and tau-lepton events that pass the search criteria, and their sum is comparable to the number of $\nu_e + \bar{\nu}_e$ cascades in each bin.  Including these would increase the Poisson probability of detecting two or more events by a factor of $\sim 2^2 = 4$.

In addition, there is a third problem, that the expected number of all events -- cascades, muon tracks, and tau leptons -- at EHE energies is large enough that some events might have been seen, but none were~\cite{Abbasi:2011ji, Aartsen:2013bka}.  The normalizations of these representative models are based on measured gamma-ray and cosmic-ray data~\cite{Takami:2007pp, Ahlers:2010fw}.  If we arbitrarily increased the normalization to increase the yields in the PeV range, that would cause an unacceptable increase in the expected number of events in the EeV range.  Cosmogenic neutrinos are thus very unlikely to be the source of the PeV events.  If they are, IceCube should quickly discover new events at higher energies.


\subsection{Atmospheric prompt neutrinos: disfavored}

Collisions of cosmic rays with atmospheric nuclei produce many unstable hadrons; these are dominantly pions, with a small fraction of kaons, and a very small fraction of mesons and baryons with heavy quarks such as charm~\cite{Gaisser:1990vg}.  The decays of many of these hadrons produce atmospheric neutrinos and muons.  Where the energy losses of these hadrons due to hadronic scattering before decay can be neglected, their spectrum and that of their daughter neutrinos follows the spectrum of the cosmic rays; otherwise, those spectra fall more steeply.

At the lowest energies, neutrinos from pions dominate.  As the energy increases, pions have increasing losses and then neutrinos from kaons dominate.  Together, these are the atmospheric conventional neutrinos.  As the energy increases further, kaons have increasing losses and then neutrinos from the decays of heavy-quark states dominate.  For these states, the decays are quite rapid, so the effects of hadron energy losses in the atmosphere are much less.  These are the atmospheric prompt neutrinos.  The conventional neutrinos have a strong zenith-angle dependence, due to the varying depth of atmosphere, but prompt neutrinos are closer to isotropic~\cite{Enberg:2008te}.

Atmospheric neutrinos have been detected with energies up to a few hundred TeV~\cite{Abbasi:2010ie}.  The spectra are consistent with atmospheric conventional neutrinos, with no prompt component identified yet.  Precise prediction of the atmospheric prompt fluxes is difficult because of uncertainties in the hadronic physics and the nuclear composition of the cosmic rays~\cite{Volkova:1980sw, Gondolo:1995fq, Battistoni:1995yv, Pasquali:1998ji, Gelmini:1999ve, Gelmini:1999xq, Martin:2003us, Candia:2003ay, Berghaus:2007hp, Enberg:2008te, Sinegovskaya:2013iaa}.

One generic prediction is that the prompt component will begin to dominate the conventional component at some high energy, due to its harder spectrum.  Another generic prediction is that the $\nu_e + \bar{\nu}_e$ flux is the same as the $\nu_\mu + \bar{\nu}_\mu$ flux for the prompt component; it is suppressed for the conventional component because pions and kaons decay primarily to muons, which are stopped in Earth before they decay.  This means that the prompt $\nu_e + \bar{\nu}_e$ component should emerge from the conventional component at lower energies than the prompt $\nu_\mu + \bar{\nu}_\mu$ component, which gives an advantage to cascade searches over track searches, despite the long range of muons, as emphasized in Ref.~\cite{Beacom:2004jb}. 

We adopt the Enberg (std.)~model~\cite{Enberg:2008te} for the atmospheric prompt neutrino flux; the components are shown in Fig.~\ref{fig:FluxOverview}.  This calculation is based on the dipole formalism in a perturbative QCD framework, which provides a way to treat gluon saturation effects at low $x$, and it assumes that the cosmic rays are protons.

There is uncertainty in the hadronic interactions, due to the extrapolation of the gluon distribution function to low $x$, and more experimental data from the LHC are needed~\cite{Itow:2011zz, Albacete:2012rx, Roland:2012dc}.  Although other perturbative QCD models may give similar results, e.g., the flux in Ref.~\cite{Martin:2003us} is about a factor of 2 below that of Ref.~\cite{Enberg:2008te}, phenomenological non-perturbative QCD approaches typically predict higher fluxes by a factor of $\sim 3-10$~\cite{Volkova:1980sw, Naumov:1998vi, Bugaev:1998bi}.  The most extreme models are already ruled out or disfavored by neutrino data~\cite{Zas:1992ci, Bugaev:1998bi, Schukraft:2013ya}.

For the atmospheric prompt fluxes, the $\nu_e + \bar{\nu}_e$ cascade spectra are shown in Fig.~\ref{fig:Cascades} and the numbers of events are given in Table~\ref{tab:presentyields}.  The slope is reasonable, in that energies near the threshold at 1 PeV are favored.  The expected number of atmospheric prompt events is $\sim 0.02$ in each of the two bins (including muon tracks and tau leptons would increase these by $\sim 50\%$, matching Ref.~\cite{Aartsen:2013bka}), so the probability of detecting at least two events is thus $\sim 10^{-4}$.  An additional problem is that the cosmic ray spectrum steepens at the knee, reducing the prompt neutrino flux~\cite{Gaisser:2013ira}.

However, the normalization of the prompt flux could easily be larger, given the substantial hadronic uncertainties, without conflicting with the neutrino measurements (which have large uncertainties) shown in Fig.~\ref{fig:FluxAstro}.  According to Refs.~\cite{Schukraft:2013ya, Aartsen:2013bka}, the normalization could be about 4 times larger; that would improve the probability by a factor of $\sim 4^2 = 16$, but it would still be very small.

The atmospheric prompt neutrino flux near 1 PeV would be even smaller if cosmic rays at higher energies are nuclei, as argued in, e.g, Refs.~\cite{Bluemer:2009zf, Gaisser:2013ira}, instead of protons, as assumed here.  The neutrino number flux per logarithmic energy bin depends on the same for the cosmic rays, which falls as $E d\Phi/dE \sim E^{1 - \gamma}$, where $\gamma \simeq 2.7$.  If cosmic rays are protons, this spectrum is used directly.  If cosmic rays are nuclei of mass number $A$, then the nucleon spectrum must be derived first.  To give the same range of nucleon energy, cosmic ray nuclei must have energies $A$ times larger, which gives a suppression $A^{1 - \gamma}$.  Taking into account the greater multiplicity of nucleons, the net suppression of the neutrino flux is $A^{2 - \gamma} \simeq A^{-0.7}$.  Therefore, if the initiating cosmic rays are dominantly nuclei, then it is even more unlikely that prompt neutrinos can explain the two observed events.

We emphasize that the atmospheric prompt neutrino hypothesis for the observed events, although disfavored, would not require the first discovery of high-energy astrophysical neutrinos.  The prompt neutrino flux has never been experimentally identified, and the theoretical uncertainties are quite large, so a very high standard must be met to reject this hypothesis.  On the other hand, if it were confirmed to be the source of the events, that would provide important and constraining information about both low-$x$ QCD and the composition of the cosmic rays.  IceCube can test the normalization of the prompt flux using both neutrinos and muons~\cite{Gelmini:2002sw, Beacom:2004jb, Gandhi:2005at, Desiati:2010wt, Gaisser:2013ira}.  The IceTop detector can reject downgoing prompt neutrinos by detecting accompanying cascades~\cite{Ishihara:2012nu,Schonert:2008is}.


\subsection{Astrophysical neutrinos: plausible}

Neutrinos are inevitably produced by cosmic-ray interactions with matter and radiation in astrophysical sources.  Many sources that may have large neutrino fluxes have been proposed, e.g., jets~\cite{Mannheim:1995mm, Halzen:1997hw, Atoyan:2001ey, Muecke:2002bi, Anchordoqui:2007tn} and cores~\cite{Stecker:1991vm, AlvarezMuniz:2004uz} of active galactic nuclei, the prompt~\cite{Waxman:1997ti, Dermer:2003zv, Murase:2005hy} and afterglow~\cite{Waxman:1999ai, Dermer:2000yd, Murase:2006dr} phases of gamma-ray bursts, newly-born neutron stars~\cite{Murase:2009pg}, early supernovae~\cite{Murase:2010cu, Katz:2011zx}, starburst galaxies~\cite{Loeb:2006tw, Thompson:2006np}, and large-scale structures and galaxy clusters~\cite{Berezinsky:1996wx, Murase:2008yt, Kotera:2009ms}.  There is a wide variety of models, each with some parameters, so roughly measuring a flux and spectrum may not identify the source.

To survey possible astrophysical diffuse sources, we consider power-law neutrino spectra, $d\Phi/dE \propto E^{-s}$.  We assume flavor ratios of $\nu_e:\nu_\mu:\nu_\tau = 1:1:1$ for neutrinos and antineutrinos, and equal fluxes of each.  (Testing flavor ratios will be important~\cite{Beacom:2003nh, Kashti:2005qa, Vissani:2013iga}.)  Because our focus is a narrow range near 1 PeV, more general spectra may be fairly characterized by power laws, and we define three cases: $s = 2$, $s = 2.5$ and $s = 3$.  The observation of two events near threshold at 1 PeV and none at higher energies strongly favors neutrino spectra that lead to adequately falling cascade spectra $E dN/dE$ beyond 1 PeV.  Below, we discuss spectra that are more general than these unbroken power laws.

We define the flux normalizations by using the largest power-law fluxes that do not exceed the measured atmospheric neutrino data at any energy, as shown in Fig.~\ref{fig:FluxAstro}.  For $s = 2$, the flux normalization for $\nu + \bar{\nu}$ in one flavor is $E^2 d\Phi/dE \simeq 0.7 \times 10^{-8}$ GeV cm$^{-2}$ s$^{-1}$ sr$^{-1}$.  This is consistent with upper bounds from IceCube~\cite{Abbasi:2011jx, Abbasi:2011ji, Abbasi:2012cu}, and is smaller than the upper range of the Waxman-Bahcall bound, $E^2 d\Phi/dE \simeq (0.3 - 1.5) \times 10^{-8}$ GeV cm$^{-2}$ s$^{-1}$ sr$^{-1}$~\cite{Waxman:1998yy}.  For $s = 2.5$, the normalization (at 1 PeV) is $E^2 d\Phi/dE \simeq 0.4 \times 10^{-8}$ GeV cm$^{-2}$ s$^{-1}$ sr$^{-1}$.  For $s = 3$, the normalization (at 1 PeV) is $E^2 d\Phi/dE \simeq 0.2 \times 10^{-8}$ GeV cm$^{-2}$ s$^{-1}$ sr$^{-1}$.  These latter two are comparable to or smaller than the nucleus-survival bound~\cite{Murase:2010gj}.

The $\nu_e + \bar{\nu}_e$ cascade spectra are shown in Fig.~\ref{fig:Cascades} and the numbers of events are given in Table~\ref{tab:presentyields}.  In the results for an ideal detector, both the slopes and normalizations of the cascade spectra are favorable, in that the cascade spectra peak near threshold at 1 PeV and reasonable numbers of events are expected.  However, in the calculation using the effective area from Ref.~\cite{Aartsen:2013bka}, the effect of the cuts on the efficiency near 1 PeV is very significant, driving down the total number of events and suppressing the importance of the first bin.  This makes the second bin, and the Glashow resonance there, much more important; for the power-law spectra, there are comparable numbers of events in the continuum and in the excess due to the resonance.  Beyond 10 PeV, the detector efficiency approaches the ideal case and, for all but the cosmogenic models, the cascade spectra are falling and the expected numbers of events are small.

For the different $s$ values in the realistic case, the total numbers of expected events might be reasonable, especially if some things are taken into account.  The normalizations for the spectra chosen in Fig.~\ref{fig:FluxAstro} could plausibly be increased by a factor of 2.  Comparable numbers of $\nu_\mu + \bar{\nu}_\mu$ and $\nu_\tau + \bar{\nu}_\tau$ CC events should be included to match the IceCube search criteria, and their sum is comparable to the number of $\nu_e + \bar{\nu}_e$ cascades in each bin.  In the $E^{-2}$ case, $\nu_\mu + \bar{\nu}_\mu$ and $\nu_\tau + \bar{\nu}_\tau$ NC events in the second bin could contribute $\sim 0.4$ events to the 0.2 CC events in the first bin.  Where the Poisson expectation is small, changing the normalization by a factor $f$ changes the probability of getting two or more events by $\sim f^2$.  The distribution of events is a larger problem: instead of favoring the lowest energies, near threshold, these cascade spectra favor higher energies in all cases.

The astrophysical models considered here are not in obvious agreement with observations, but this depends on the details of the  efficiency near threshold, so we must withhold judgment until there are results from new searches.  It is plausible that astrophysical scenarios could explain the observed events.  Taking the large uncertainties into account, spectra less steep than $E^{-2}$ seem to be disfavored by the spectrum shape, and spectra more steep than $E^{-3}$ seem to be strongly disfavored by the spectrum normalization.  The most important thing is to improve the efficiency at energies below 1 PeV, where the number of events might be much larger.


\subsection{What conclusions can we draw now?}

None of the sources above immediately fits the key observed properties of the data: two cascade events, very close in energy to each other and the analysis threshold, no cascades at higher energies, and no other types of events.  How can this be?  We focus on steady diffuse fluxes here and then mention other possibilities below.

One possibility is improbable fluctuations.  These two events might be caused by astrophysical neutrino signals, and what was seen was a lucky fluctuation.  Reconciling what was and was not seen may be challenging.  Or these two events might be caused by atmospheric neutrino or muon backgrounds, and what was seen was an unlucky fluctuation.  With the expected rates, this is very unlikely; further study is needed to be sure there are no surprises with muon-induced backgrounds.

Another possibility, which we think is unlikely, is that the effective area or the relation between the number of detected photoelectrons and cascade energy is not completely understood.  The search strategy was optimized for cosmogenic neutrinos in the EeV range, and perhaps there are subtleties near 1 PeV, the edge of their range~\cite{Aartsen:2013bka}.  The IceCube Collaboration takes great care in their analyses and papers, but the possibility of some revisions being needed must be considered because of the seeming paradox of detecting two events near threshold, where the efficiency is only $\sim 20\%$.

The last possibility is that these are astrophysical neutrinos, but that the spectrum is peaked.  If the flavor ratios are near unity, as expected, this would require some fine-tuning of the spectrum.  Figure~\ref{fig:FluxAstro} shows that there are strong upper limits on the flux at a few hundred TeV to avoid conflict with atmospheric conventional neutrino data, and Fig.~\ref{fig:Cascades} shows that should also be strong upper limits on the flux at several PeV to avoid conflict with the non-observation of events where the detection efficiency is favorable.  In the decade in energy in between, the flux should be large enough to make it probable to detect two events despite the low efficiency near 1 PeV.  Some examples of peaked spectra include gamma-ray bursts~\cite{Cholis:2012kq, Liu:2012pf}, very heavy dark matter decay~\cite{Murase:2012xs, Feldstein:2013kka}, and cosmic ray proton interactions~\cite{Essey:2009ju, Essey:2010er, Kalashev:2013vba}.

\begin{figure}[t]
\includegraphics[width=\columnwidth]{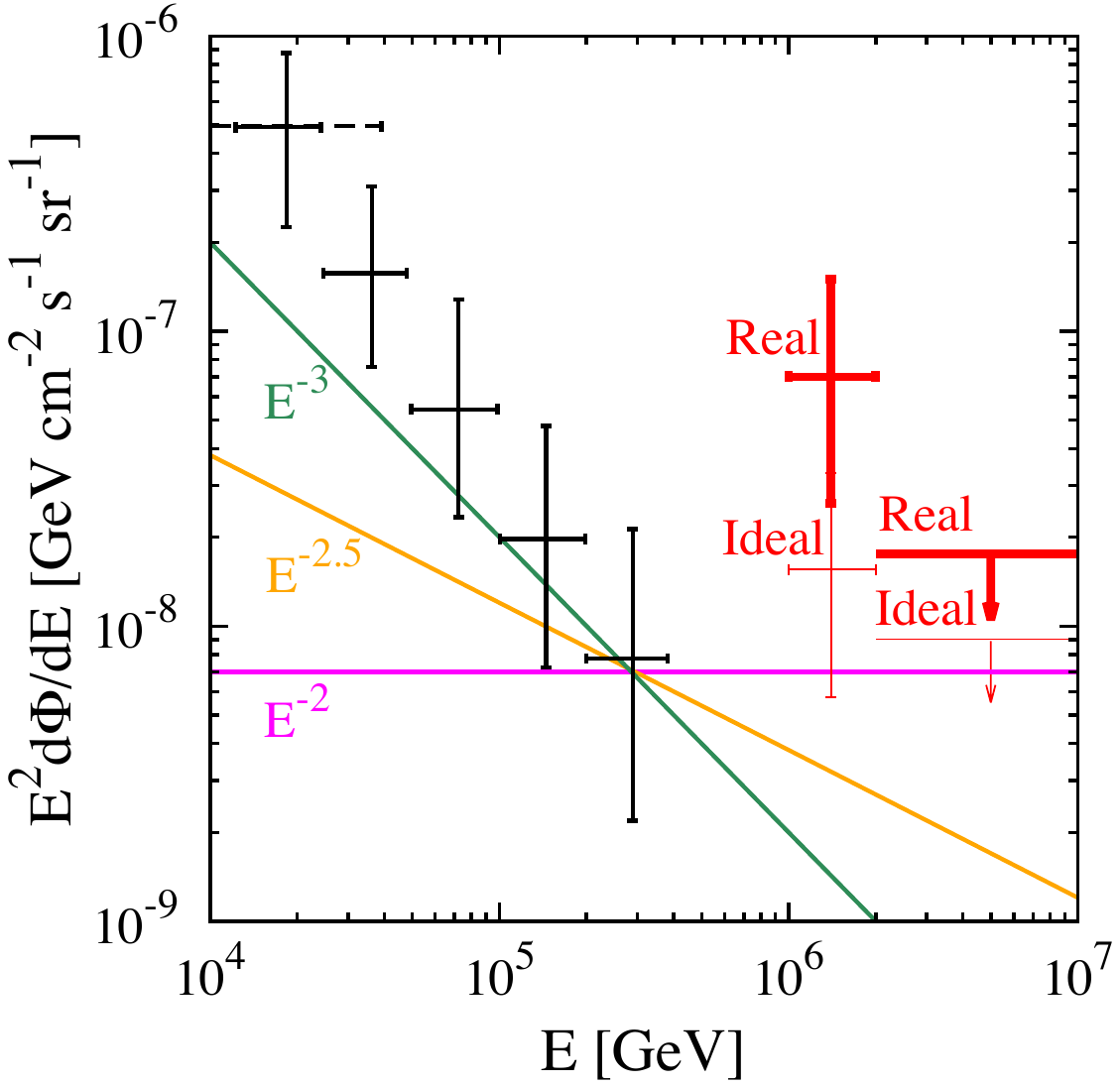}
\caption{Example neutrino fluxes, as in Fig.~\ref{fig:FluxAstro}, for one flavor of $\nu + \bar{\nu}$, assuming equal flavor ratios.  In the 1--2 PeV and 2--10 PeV bins, we show our estimates of the flux normalization required to match the observations of two events and zero events, respectively, for an $E^{-2}$ spectrum in each bin separately.  We show the 68\% confidence-level uncertainty range for the first bin and the 90\% confidence-level upper limit for the second~\cite{Feldman:1997qc}.  The ``Real" case uses the right panel of Fig.~\ref{fig:Cascades} (based on Ref.~\cite{Aartsen:2013bka}), while the ``Ideal" case uses the left.}
\label{fig:FluxConstraints}
\end{figure}

We highlight these constraints on astrophysical neutrino spectra in Fig.~\ref{fig:FluxConstraints}, which focuses on the most important region of Fig.~\ref{fig:FluxAstro}.  We show the normalizations of an $E^{-2}$ spectrum in the three energy ranges separately, set by Fig.~\ref{fig:FluxAstro}, the observation (and hence expectation) of two events in the first bin, and the observation of zero events in the second bin, respectively.  (We always quote neutrino fluxes for one flavor of $\nu + \bar{\nu}$, assuming equal flavor ratios, whereas some authors quote the sum of all three flavors.)  These results suggest a break in the spectrum at several hundred TeV and another break or cutoff at about 2 PeV.  For a different spectrum shape or choice of bins, these constraints would change.  Still, the nominal conflicts between fluxes in different energy ranges are startling, and indicate tensions that need to be resolved.

The dominant uncertainties are those shown in Fig.~\ref{fig:FluxConstraints}.  We fix the power-law normalizations in Fig.~\ref{fig:FluxAstro} by demanding that they not exceed the measured points.  This leaves no room for the expected atmospheric conventional neutrinos, but the uncertainties are large, probably even larger than the quoted factors of a few up or down.  The Poisson uncertainties on the fluxes in our two bins are significant.  Our calculations of the expected numbers of events are reasonably precise, though we make approximations throughout at the level of a few tens of percent.  These include the form of the event rate equations, approximating the $d\sigma/dy$ distributions and Earth attenuation, and neglecting the small numbers of expected events below 1 PeV and above 10 PeV.

If the true spectrum is not peaked, then the most likely scenario is that there should be an excess in the low-energy muon neutrino data (now seen in Ref.~\cite{Schukraft:2013ya}), that the observation of the two PeV events was a fortunate upward fluctuation, and that there should be a cutoff at about 2 PeV.  In this case, our results show that the preferred power-law spectrum is around $E^{-2}$.  The strong constraint on an astrophysical neutrino flux shown in Fig.~\ref{fig:FluxOverview}, $E^2 d\Phi/dE < 0.9 \times 10^{-8}$ GeV cm$^{-2}$ s$^{-1}$ sr$^{-1}$~\cite{Abbasi:2011ji}, would apply to an $E^{-2}$ spectrum that held over the full energy range shown there.  See also the preliminary differential constraints shown in Ref.~\cite{Ishihara:2012nu}.

\section{Future neutrino observations}

As we show above, the source of the two cascade events in IceCube remains unknown, though some possibilities can already be excluded.  With such a small sample and such large uncertainties, it is not yet possible to make very precise statements.  We now show that analyses of existing cascade data at lower energies have great potential to quickly reveal the source of these events.  Searches for muon tracks in IceCube are quite mature, with atmospheric neutrino events measured up to a few hundred TeV~\cite{Abbasi:2010ie}.  To measure the smaller fluxes at higher energies, greater exposure is needed, which will simply take time.  In contrast, searches for cascades with measured atmospheric neutrino events are relatively recent and the spectra only go up to 10 TeV~\cite{Aartsen:2012uu}.

A comprehensive exploration below 1 PeV, where there might be many more events, is needed in both the track and cascade channels.  In the following, we first review muon track detection in IceCube.  Cascade detection is discussed in detail above.  Here, one important difference is that $\nu_\tau + \bar{\nu}_\tau$ CC events are now included as cascades for the astrophysical scenarios (but not for atmospheric prompt neutrinos, which have a small $\nu_\tau + \bar{\nu}_\tau$ flux) because the tau-lepton track length below 1 PeV is short.  We show how our new results on the predicted spectra can differentiate between possible scenarios.

The following is for the ideal case or ``theorist's approach," because the detailed properties of IceCube for future searches are not yet known, as new strategies to isolate signals from backgrounds will be developed.  The true efficiency will be somewhat less, e.g., due to cuts to reject backgrounds and because outward-directed signal events near the surface will not deposit enough energy.  In addition, the spectrum shapes will suffer some smearing due to energy resolution.  The most important point of realism that we do include is that we plot our results in terms of measurable energy, not neutrino energy, as this gives better separation of signals and backgrounds.


\subsection{Muon tracks in IceCube}

Muons are produced by the CC interactions of $\nu_\mu + \bar{\nu}_\mu$ with nucleons~\cite{Gandhi:1998ri, Connolly:2011vc, CooperSarkar:2011pa, Block:2013nia}.  The initial muon energy is $E_\mu \simeq (1 - \langle y \rangle) E_\nu \simeq 0.75 E_\nu$ for $E_\nu \sim 1$ PeV~\cite{Gandhi:1995tf}.  Because of their small energy loss rate and long lifetime, muons produce long tracks; above 1 PeV, the muon range in ice is $\sim 15$ km and varies logarithmically with energy.  Those produced inside IceCube are contained-vertex muons, whereas those produced outside are through-going muons.    For contained-vertex muons, the hadronic energy will be deposited in the detector, while it is lost for through-going muons.

We present our results in terms of the energy of the muon as it first appears in the detector, due to being created there or when it first enters.  This is measurable and provides the most information about the neutrino spectrum~\cite{Beacom:2003nh, Kistler:2006hp, Cholis:2012kq}.  The average muon energy loss rate is $-dE_\mu/dx = \alpha + \beta E_\mu$~\cite{Lipari:1991ut, Dutta:2000hh}.  In the TeV range and above, the radiative term ($\beta E_\mu$) dominates the ionization ($\alpha$) term.  We take $\alpha \simeq 2 \times 10^{-3}$ GeV cm$^2$ g$^{-1}$ (its low-energy value) and $\beta \simeq 5 \times 10^{-6}$ cm$^2$ g$^{-1}$ (near 1 PeV).  The muon energy can be measured by the fluctuations in its radiative losses, and a precision of a factor of 2 is expected~\cite{Whitehorn:tevpa2012}.  The present EHE search simply measures the number of detected photoelectrons produced by an event, which utilizes less information.

The complete measurable muon spectrum is
\begin{equation}
\left(\frac{dN}{dE_\mu}\right)_{tracks} = \left(\frac{dN}{dE_\mu}\right)_{cont} + \left(\frac{dN}{dE_\mu}\right)_{thru},
\label{eq:tracks}
\end{equation}
where the same value of $E_\mu$ comes from different ranges of neutrino energy in the two cases.  For simplicity, we add these event classes, though they should be separable.  In the following, through-going events are about 3 times more numerous than contained-vertex events for an $E^{-2}$ spectrum, and about 1.5 times more so for an $E^{-3}$ spectrum.  We consider only upgoing neutrino-induced muons, to avoid the large backgrounds from downgoing atmospheric muons.  In principle, it should be possible to include some downgoing contained-vertex events~\cite{Schonert:2008is}.

The muon spectrum from contained-vertex events~\cite{Gaisser:1990vg, Kistler:2006hp} is similar to that for electron cascades and is
\begin{eqnarray}
\left(\frac{dN}{dE_\mu}\right)_{cont}
& & \simeq  2\pi \, \rho \, N_A \, V \, T \\
& \times &  \int^{0}_{-1} d(\cos{\theta_z}) \, \dfrac{d\Phi}{dE_\nu}(E_\nu) \, \sigma(E_\nu) \, e^{-\tau(E_\nu, \cos{\theta_z})}\,. \nonumber
\label{eq:contained muon}
\end{eqnarray}
Here we assume $E_\mu \simeq E_\nu$ because the hadronic cascade will contribute to the energy deposited.

The muon spectrum from through-going events~\cite{Gaisser:1990vg, Kistler:2006hp}, taking into account the increase in the effective volume of the detector due to the long muon range, is
\begin{eqnarray}
\left(\frac{dN}{dE_\mu}\right)_{thru}
& & \simeq 2\pi \, \rho \, N_A \, A \, T \\
& \times & \int^{0}_{-1} d(\cos{\theta_z}) \,
\dfrac{1}{\rho\,(\alpha+\beta E_{\mu})} \nonumber \\
& \times & \int_{E_\mu}^{E_{\rm high}} dE_{i} \,\dfrac{d\Phi}{dE_i}(E_i) \, \sigma(E_i) \, e^{-\tau(E_i, \cos{\theta_z})} \,, \nonumber
\label{eq:muon rate}
\end{eqnarray}
where $E_i$ is the initial neutrino energy and $E_{high}$ its maximum value, which depends on the distance to the horizon at that zenith angle; for upgoing events, $E_{high}$ is effectively infinite.  Instead of the detector volume, the detector area $A \simeq 1$ km$^2$ and a term reflecting the muon range appear.  We neglect the large fluctuations in the muon energy-loss rate~\cite{Lipari:1991ut, Dutta:2000hh}.  This and the preceding event rate equations also neglect the integration over $d\sigma/dy$, which can affect the results by a few tens of percent, which is within our uncertainties.


\begin{figure*}[t]
\centering
\includegraphics[width=\columnwidth]{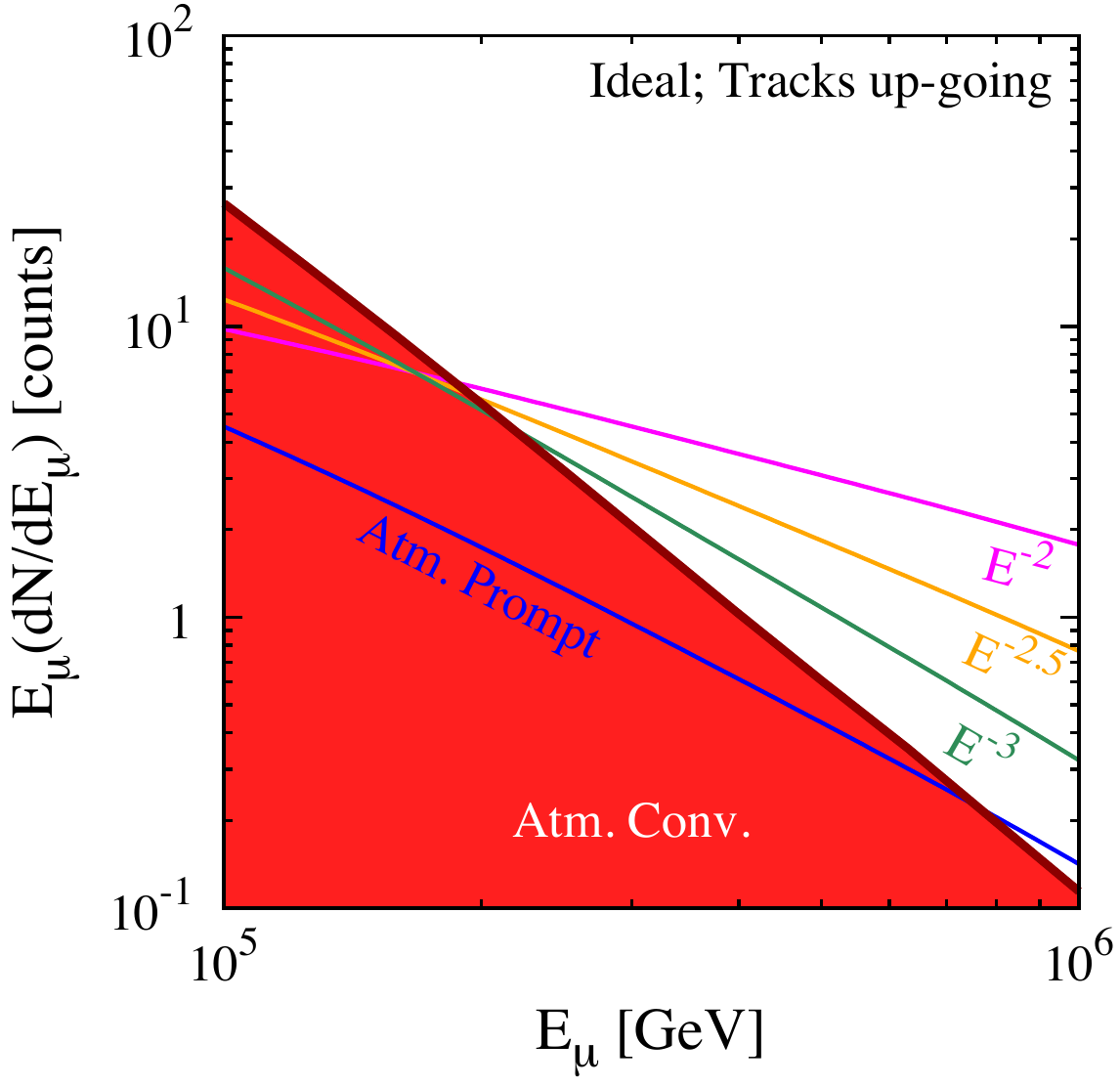}
\hspace{0.25cm}
\includegraphics[width=\columnwidth]{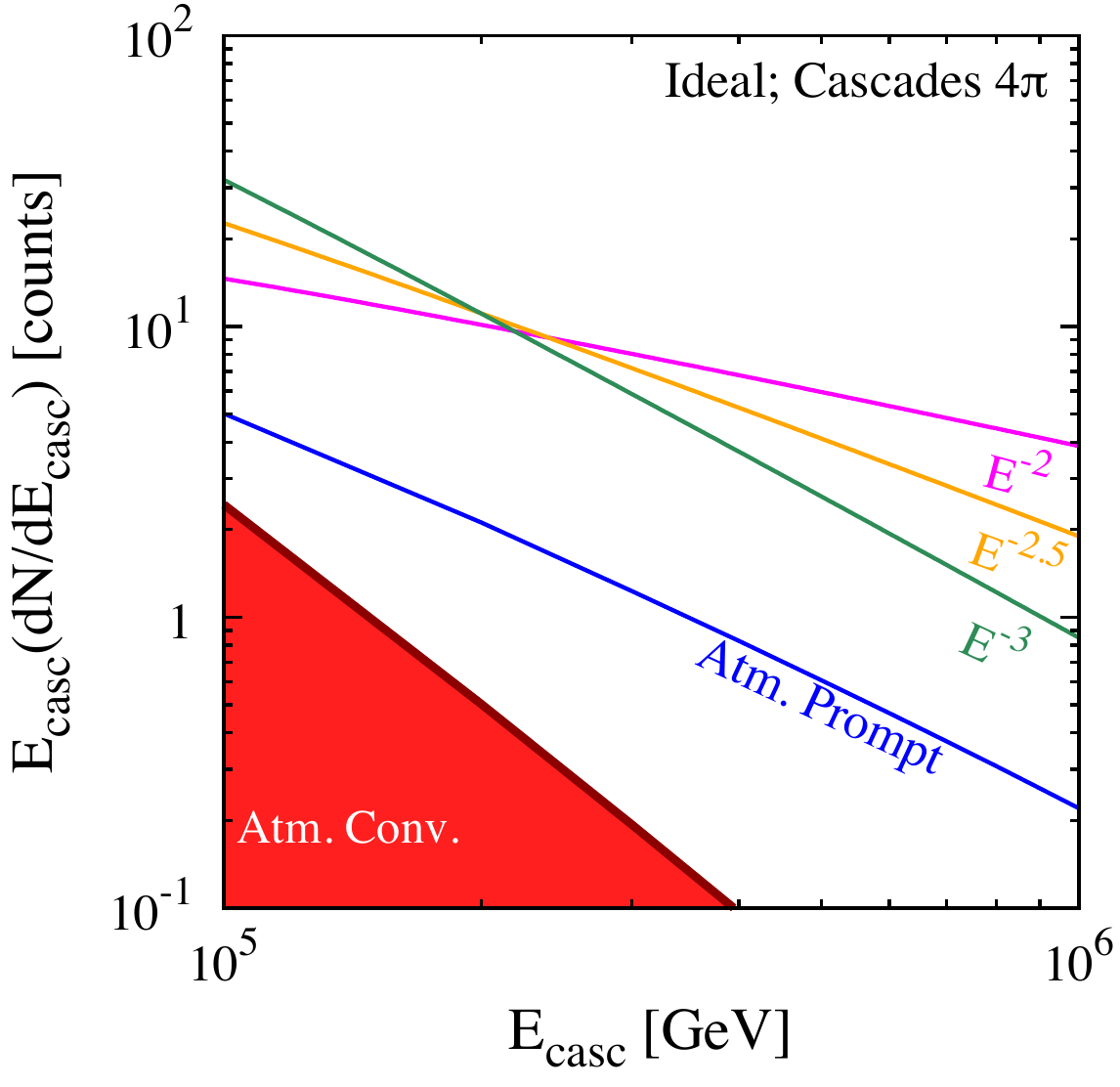}
\caption{Predictions for measurable spectra in two years of the full IceCube for various neutrino spectra considered above.  ({\bf Left Panel}) $E dN/dE$ for neutrino-induced muons (upgoing only), where the muon energy is measured as it first appears in the detector, whether as a contained-vertex or through-going event.  ({\bf Right Panel}) $E dN/dE$ for neutrino-induced cascades (all directions), where the cascade energy is measured as deposited in the detector, whether as a CC or NC event.  As above, the number of events in a region is proportional to the integrated area, i.e., to the height times the logarithmic energy range.}
\label{fig:Future}
\end{figure*}

\subsection{Predicted spectra below 1 PeV}

Figure~\ref{fig:Future} shows our predicted track and cascade spectra for two years of the full IceCube; the numbers of events are given in Table~\ref{tab:futureyields}.  It is likely that much of this exposure time can be obtained from existing data with new analyses targeted to this energy range.   All input neutrino fluxes are normalized as in previous figures.  To avoid over-extrapolating the power-law astrophysical fluxes and to focus on the energy range with the best ratio of signal to background, we show results only down to 0.1 PeV, though IceCube should go to lower energies.

The left panel shows that analyses with muon tracks are limited by the large atmospheric conventional background, so that the astrophysical signals will only emerge above a few hundred TeV, especially once the smearing effects of energy resolution are taken into account.  Even if just contained-vertex muons are selected, the background due to atmospheric conventional $\nu_\mu + \bar{\nu}_\mu$ will be dominant until high energies, where the statistics are low.  There is now some excess at the highest energies in the IceCube neutrino-induced muon data~\cite{Schukraft:2013ya}.  However, it is difficult to judge the significance when the results have been processed by unfolding to estimate the spectrum in terms of neutrino energy, which mixes different ranges of measurable muon energy and gives strongly correlated uncertainties.  When spectra are shown in terms of muon energy, there is better separation of signal and background and then even a small number of signal events at high energy can be quite significant~\cite{Kistler:2006hp}.

\begin{table}[b]
\caption{Expected numbers of track and cascade events (ideal case or ``theorist's approach"), obtained by integrating the curves in each panel of Fig.~\ref{fig:Future} over the range 0.1--1 PeV.}
\setlength{\extrarowheight}{4pt} 
\begin{ruledtabular}
\begin{spacing}{1.1}
\begin{tabular}{lcc}
Possible Source & ${\rm N_{track}}$ & ${\rm N_{casc}}$ \\ 
\hline
Atm. Conv.~\cite{Abbasi:2011jx, Honda:2006qj} & 11 & 1 \\
\hline
Atm. Prompt~\cite{Enberg:2008te} & 3 & 4 \\
\hline
Astrophysical $E^{-2}$ & 11 & 19 \\
Astrophysical $E^{-2.5}$ & 10 & 20 \\
Astrophysical $E^{-3}$ & 9 & 20 \\
\end{tabular}
\end{spacing}
\end{ruledtabular}
\label{tab:futureyields}
\end{table}

The right panel shows that the prospects for cascades are extremely promising, because the atmospheric conventional background is strongly suppressed, as first shown in Ref.~\cite{Beacom:2004jb}.  The difference in cascade rates at 1 PeV seen between the left panel of Fig.~\ref{fig:Cascades} and the right panel of Fig.~\ref{fig:Future} is due to the latter including $\nu_\tau + \bar{\nu}_\tau$ events (factor of 2), the slightly different exposure times, and the former including energy resolution smearing.

Even if the efficiency is reduced from that shown in Fig.~\ref{fig:Future}, it should still be possible to detect potentially large numbers of cascade events with minimal backgrounds.  This could quickly discover an astrophysical flux.  The atmospheric conventional neutrinos and even the atmospheric prompt neutrinos are negligible backgrounds.  The cascade spectrum shape will be a powerful diagnostic of the neutrino spectrum shape, because $E_{casc} \simeq E_\nu$ for the dominant CC events and good energy resolution for cascades. The normalizations of these spectra are the largest values that do not conflict with the measured atmospheric neutrino data shown in Fig.~\ref{fig:FluxAstro}.  If the normalizations were instead set by the requirement of producing the two PeV events, then the curves in Fig.~\ref{fig:Future} would cross near 1 PeV and the differences between them would be much larger below 1 PeV.

Even though there are essentially no neutrino-induced backgrounds for cascade signals, there may be backgrounds induced by downgoing atmospheric muons~\cite{Aartsen:2013bka}.  The cascade analysis that measured the conventional atmospheric neutrino spectrum up to 10 TeV, as shown in Fig.~\ref{fig:FluxAstro}, used the small inner DeepCore detector as the active volume and the rest of IceCube as a veto~\cite{Aartsen:2012uu}.  It should be possible to extend this idea as a function of energy, effecting a series of nested inner detectors and outer veto layers, with larger inner volumes than DeepCore probing the smaller fluxes at higher energies.


\section{Conclusions}


\subsection{Summary and Outlook}

The observation of two cascade events near 1 PeV~\cite{Ishihara:2012nu, Aartsen:2013bka} is a remarkable achievement that follows more than two decades of pioneering work by the AMANDA and IceCube Collaborations~\cite{Barwick:1991ur, Lowder:1991uy, Andres:1999hm, Ahrens:2003ix, Achterberg:2006md}.  It is very likely that these are neutrino-induced events, possibly the first high-energy astrophysical neutrinos ever observed, opening a new era.  A high burden of proof will be needed to reject all hypotheses based on a terrestrial origin and to accept any based on an astrophysical origin.

We provide a comprehensive general study of these PeV events and their possible origin as a diffuse flux~\cite{LahaAspen:2013, LahaIPA:2013}.  We apply physical insights to characterize the nature of the events and to define the framework for analyzing possible source spectra.  We systematically analyze several possible neutrino sources and backgrounds and draw conclusions about whether they can explain the observed events in light of realistic detector modeling and other constraints.  We show how IceCube can most quickly uncover the nature of these events  with searches at lower energies, for which we make detailed predictions.

The search efficiency near the analysis threshold at 1 PeV is $\sim 20\%$, which makes it surprising that two events were observed there.  As shown in Fig.~\ref{fig:FluxConstraints}, a high neutrino flux near 1 PeV is needed to counteract this low efficiency, while low fluxes are needed at slightly lower and higher energies to avoid overproducing events there.  A relatively narrow spectrum peak might be called for~\cite{Cholis:2012kq, Liu:2012pf, Essey:2009ju, Essey:2010er, Kalashev:2013vba, Murase:2012xs, Feldstein:2013kka}.  On the other hand, besides the significant uncertainties shown in Fig.~\ref{fig:FluxConstraints}, the details depend on the efficiency where it is small and changing rapidly.

Some possible neutrino sources are already quite disfavored in any case, as shown in Fig.~\ref{fig:Cascades}.  For atmospheric conventional neutrinos, the expected rates are far too small.  The cascade backgrounds induced by atmospheric muons also seem to be too small~\cite{Aartsen:2013bka}.  Atmospheric prompt neutrinos are also disfavored, though special caution is needed because this source is guaranteed, has never been identified experimentally, and has large theoretical uncertainties.  For cosmogenic fluxes (those produced in the propagation of ultra-high-energy cosmic rays), the expected rate is too small and the cascade spectrum increases with energy, contrary to observations.

We also consider a steady diffuse background of neutrinos produced in astrophysical sources, parameterizing these as power-law spectra for energies near 1 PeV, and assuming equal flavor ratios.  Power-law spectra between $E^{-2}$ and $E^{-3}$ are plausible, with $E^{-2}$ (with a cutoff at about 2 PeV) being the most likely.  There are tensions regarding the normalization and slope of such models, but these are subject to the above uncertainties.

The most important thing for IceCube to do is to improve the efficiency of searches at and below 1 PeV.  We show in detail, including in Fig.~\ref{fig:Future}, how such searches can differentiate between possible scenarios for the observed PeV events.  Even in the absence of one or both of these events, there is tremendous discovery potential for cascade searches in this energy range.  The detection of cascade events has long been recognized as important, as a probe of $\nu_e + \bar{\nu}_e$ and because the good fidelity between cascade and neutrino energy allows reconstruction of the neutrino spectrum~\cite{IceCube:2001}.  As first shown by Beacom and Candia~\cite{Beacom:2004jb} in 2004, there is a strong suppression of the atmospheric conventional neutrino background for the cascade channel relative to the muon track channel, giving improved sensitivity at lower energies.

Our results on cascades go well beyond those in Ref.~\cite{Beacom:2004jb} and will be generally useful for future searches.  In addition to adopting updated fluxes, we provide a detailed discussion of the effects of many realistic IceCube detector properties.  We show how to best display and interpret cascade spectra over a wide energy range, including near the Glashow resonance, where energy resolution effects must be included.  We compare cascade and track spectra, with both presented in terms of detectable energy instead of neutrino energy.

Many of our considerations would easily carry over for point sources or collections thereof.  For the same two PeV events detected, which sets the total flux required, point sources would be easier to separate from the conventional atmospheric neutrino background because the relevant solid angle would be smaller than the full sky.

Whatever the origin of these two events, their detection is an important milestone in advancing our knowledge of the high-energy Universe, and we congratulate the IceCube Collaboration on this success.  Now that the construction of the IceCube detector is complete, neutrinos will be detected at a faster rate, and great progress is expected soon, which we eagerly await.


\subsection{Impact of new results}

As this paper was being completed (for early results, see Refs.~\cite{LahaAspen:2013, LahaIPA:2013}), IceCube announced the detection of new events~\cite{Whitehorn:IPA2013}.  These preliminary data shed light on the PeV events and seem to strengthen the case that their origin is astrophysical.  There are no serious disagreements with our results and many of our assumptions are now confirmed.  Here we summarize their most important new results and our interpretation of them.

The basic aspects of the data fit within the framework we consider.  The events are consistent with being uniform in the volume, so are likely not due to backgrounds induced by downgoing atmospheric muons.  No remarks are made about the arrival times of the events, so presumably they are consistent with being from a steady source.  The distribution of arrival directions is consistent with isotropy subject to expected attenuation in Earth, so consistent with a diffuse source.

New search criteria improved the efficiency at 1 PeV by a factor of 3 (i.e., part of the possible factor of 5 noted above); the improvements at nearby energies vary with energy.  No new events were found near 1 PeV, or at higher energies, which indicates that our choice of bins in the PeV range was reasonable and that the former observation of two events must have been a lucky fluctuation.  This follows from Fig.~\ref{fig:Cascades} and the surrounding discussion, and acts to reduce the tensions shown in Fig.~\ref{fig:FluxConstraints}.

The new criteria also provided some efficiency at energies well below the previous threshold at 1 PeV.  There are 19 new cascade events between 0.03 and 0.3 PeV.  Only six of these are above 0.1 PeV, where the atmospheric neutrino backgrounds are minimal.  The number above 0.1 PeV is reasonable for the $E^{-2}$ spectrum above.  The lack of events above 0.3 PeV supports the detection of the two PeV events being a lucky fluctuation.

The new events also include 7 contained-vertex muon events, all between 0.03 and 0.3 PeV; all but one are below 0.1 PeV.  The new search criteria still suppress $\nu_\mu + \bar{\nu}_\mu$ detection relative to $\nu_e + \bar{\nu}_e$ detection, a small fraction of track to cascade events is expected.  The left panel of our Fig.~\ref{fig:Future} shows that atmospheric conventional neutrinos dominate in this energy range, including for contained-vertex muon events.  For equal flavor ratios, cascade events are much more likely to be signals than are track events, so these events should not be mixed.

It is stated that an $E^{-2}$ spectrum is a reasonable fit, provided there is a spectrum cutoff in the PeV range, as we independently show.  The right panel of our Fig.~\ref{fig:Future} shows that the ratios of numbers of events near 0.1 PeV to those near 1 PeV are quite distinct for different power-law spectra, so this will be a powerful test of the spectrum.  More information is needed on the consistency of an $E^{-2}$ spectrum with lower-energy neutrino data of all flavors.  The difficulties we point out in Fig.~\ref{fig:FluxConstraints} would be somewhat alleviated if analyses of that data show some excesses near a few hundred TeV, as is now reported.


\subsection{Astrophysical implications}

Many models of astrophysical neutrino sources have been proposed.  There are two key requirements for viable scenarios to explain the IceCube results.  First, the cosmic-ray energy injection rate and meson production efficiency must be sufficient to give a neutrino flux of at least $E^2 d\Phi/dE \sim 10^{-8}$ GeV cm$^{-2}$ s$^{-1}$ sr$^{-1}$ near 1 PeV.  Second, since protons with energy $\varepsilon_p$ at a typical redshift $z \sim 1$ lead to neutrinos with energy $E_\nu \sim 2~{\rm PeV}~(\varepsilon_p/100~{\rm PeV})$, sources should be able to accelerate protons to energies close to the iron/second knee~\cite{Murase:2008yt}.  In addition, a break at high energies seems to be required, and the spectrum may even be peaked.

Proton-photon ($p\gamma$) interactions are dominant for PeV neutrino production in most models of active galactic nuclei (AGN) and gamma-ray bursts (GRBs)~\cite{Rachen:1998fd}.  Protons typically interact near threshold with photons of energy $\varepsilon_\gamma$, so $\varepsilon_p \varepsilon_{\gamma} \sim 0.16~{\rm GeV}^2 \, \Gamma^2$, where $\Gamma$ is the Lorentz factor.  Then $\pi^-$ production is suppressed and fewer antineutrinos are produced.  In addition, flavor ratios are affected by muon cooling in magnetized sources~\cite{Kashti:2005qa}.  In the $p\gamma$ case, the neutrino spectrum is hard at low energies and typically has a peak depending on source properties.

In AGN jet models~\cite{Mannheim:1995mm, Halzen:1997hw, Atoyan:2001ey, Muecke:2002bi, Anchordoqui:2007tn}, the neutrino spectrum peaks at $\sim 10-1000$~PeV because $\Gamma \sim 10$ and the observed photon spectra of luminous blazars peak at $\varepsilon_\gamma \sim 0.1-10$~eV.  The spectrum is expected to be rising at energies above the PeV range, as for cosmogenic models, which are disfavored.  In AGN core models~\cite{Stecker:1991vm, AlvarezMuniz:2004uz}, where neutrinos are produced not far from accretion disks, a peak in the PeV range is possible, though optimistic cases have been ruled out.

In GRB prompt emission models~\cite{Waxman:1997ti, Dermer:2003zv, Murase:2005hy}, PeV neutrinos are expected because $\varepsilon_\gamma \sim 1$~MeV and $\Gamma \sim 300$.  Due to strong meson cooling, a break or even a bump was predicted around $1-100$~PeV~\cite{Murase:2008sp, Wang:2008zm}.  Although this spectrum shape may be appealing, stacking searches by IceCube set limits of $E^2 d\Phi/dE \lesssim 0.1 \times 10^{-8}$ GeV cm$^{-2}$ s$^{-1}$ sr$^{-1}$~\cite{Abbasi:2012zw, Hummer:2011ms, He:2012tq, Liu:2012pf}, well below the required flux.  However, many transients like low-luminosity GRBs are missed; some predictions are $\sim 10$ times larger than this limit and have a peak or break in the PeV range~\cite{Murase:2006mm, Gupta:2006jm, Cholis:2012kq}.  Although neutrinos can be produced in GRB afterglows~\cite{Waxman:1999ai, Dermer:2000yd, Murase:2006dr}, their typical energy is much higher than 1 PeV, as in the AGN jet model, so explaining the IceCube PeV events is difficult.

Proton-proton ($pp$) interactions are dominant for PeV neutrino production in starburst galaxies and large-scale structures.  Many pions of all types are produced in each scattering, and the neutrino spectrum basically follows the proton spectrum~\cite{Kelner:2006tc}, with equal ratios of neutrinos and antineutrinos and of flavors after mixing.

Starburst galaxies contain many massive stars, which lead to supernovae that may produce cosmic rays.  Most of the cosmic ray power would be lost to neutrinos and gamma rays due to interactions in the high-column-density material, and detections of gamma rays from nearby galaxies~\cite{Acero:2009nb} supports this idea.  The predicted flux is $E^2 d\Phi/dE \sim (0.1 - 10) \times 10^{-8}$ GeV cm$^{-2}$ s$^{-1}$ sr$^{-1}$, with a possible cutoff~\cite{Loeb:2006tw, Thompson:2006np}, though it is uncertain if $\sim100$~PeV protons (rather than heavy nuclei) are produced in these galaxies.

Large-scale structures (especially galaxy clusters) are gigantic reservoirs of cosmic rays that may be accelerated at structure
formation shocks and supplied by contained AGN~\cite{Berezinsky:1996wx}.  PeV neutrinos are produced via $pp$ interactions with the intracluster medium.  The expected flux is $E^2 d\Phi/dE \sim (0.1 - 1) \times 10^{-8}$ GeV cm$^{-2}$ s$^{-1}$ sr$^{-1}$, and a break due to the diffusive escape or maximum energy of cosmic rays has been predicted~\cite{Murase:2008yt}.

The possible connection with extragalactic cosmic rays is intriguing, because a neutrino flux of $E^2 d\Phi/dE \sim 10^{-8}$ GeV cm$^{-2}$ s$^{-1}$ sr$^{-1}$ is comparable to the Waxman-Bahcall bound~\cite{Waxman:1998yy} derived from the ultra-high-energy cosmic ray flux.  However, PeV neutrinos correspond to protons at lower energies, near 100 PeV, and higher-energy neutrinos have not been detected, despite the increasing effective area.  If ultra-high-energy cosmic rays are heavy nuclei, as suggested by Auger, then the neutrino flux from their sources is much lower than the Waxman-Bahcall bound~\cite{Murase:2010gj}.  

To conclude our discussion of astrophysical neutrino fluxes, there is so far no obvious source that explains all aspects of the IceCube data.  Many models (e.g., GRB prompt, starburst galaxies, and large-scale structures) seem compatible with the data, though some models (e.g., AGN jets and GRB afterglow) are already disfavored.  Interestingly, the neutrino flux sensitivity is approaching that needed to probe the sources of the ultra-high-energy cosmic rays.  More experimental data and theoretical studies are needed to unravel the mysteries of the high- and ultra-high-energy universe.


\bigskip
\noindent
{\bf Acknowledgments:}
We thank Markus Ahlers, Atri Bhattacharya, Kfir Blum, Raj Gandhi, Aya Ishihara, Peter Meszaros, and especially Amy Connolly, Francis Halzen, Albrecht Karle, Kenny Ng, Carsten Rott, Nathan Whitehorn, and Shigeru Yoshida for helpful discussions.  RL also thanks the participants at the workshops where early results from this work were presented~\cite{LahaAspen:2013, LahaIPA:2013} for their helpful feedback.  RL and JFB were supported by NSF Grant PHY-1101216 to JFB, and BD, SH, and KM were supported by CCAPP during the early part of this work.


\newpage

\bibliographystyle{kp}
\interlinepenalty=10000
\tolerance=100
\bibliography{Bibliography/references}

\begingroup\raggedright\begin{thebibliography}{143}
\expandafter\ifx\csname natexlab\endcsname\relax\def\natexlab#1{#1}\fi

\bibitem[Ishihara(2012)]{Ishihara:2012nu}
A.~Ishihara, ``{IceCube:Ultra-high Energy Neutrinos}'', {\em Neutrino 2012
  Conference \url{http://neu2012.kek.jp/index.html}}, 2012.

\bibitem[Aartsen et~al.(2013)]{Aartsen:2013bka}
{\bfseries IceCube Collaboration} Collaboration, M.~Aartsen {\em et~al.},
  ``{First observation of PeV-energy neutrinos with IceCube}'', {\em
  Phys.Rev.Lett.} {\bfseries 111} (2013) 021103,
 \href{http://xxx.lanl.gov/abs/1304.5356}{ arXiv:1304.5356}.

\bibitem[Greisen(1960)]{Greisen:1960wc}
K.~Greisen, ``{Cosmic ray showers}'', {\em Ann.Rev.Nucl.Part.Sci.} {\bfseries
  10} (1960)
63--108.

\bibitem[Pontecorvo(1963)]{Pontecorvo:1963wk}
B.~Pontecorvo, ``{The neutrino and its role in astrophysics}'', {\em
  Usp.Fiz.Nauk} {\bfseries 79} (1963)
3--21.

\bibitem[Lande(1979)]{Lande:1979gr}
K.~Lande, ``{Experimental Neutrino Astrophysics}'', {\em
  Ann.Rev.Nucl.Part.Sci.} {\bfseries 29} (1979)
395--410.

\bibitem[Gaisser et~al.(1995)Gaisser, Halzen, and Stanev]{Gaisser:1994yf}
T.~K. Gaisser, F.~Halzen, and T.~Stanev, ``{Particle astrophysics with
  high-energy neutrinos}'', {\em Phys.Rept.} {\bfseries 258} (1995) 173--236,
 \href{http://xxx.lanl.gov/abs/hep-ph/9410384}{ arXiv:hep-ph/9410384}.

\bibitem[Learned and Mannheim(2000)]{Learned:2000sw}
J.~Learned and K.~Mannheim, ``{High-energy neutrino astrophysics}'', {\em
  Ann.Rev.Nucl.Part.Sci.} {\bfseries 50} (2000)
679--749.

\bibitem[Halzen and Hooper(2002)]{Halzen:2002pg}
F.~Halzen and D.~Hooper, ``{High-energy neutrino astronomy: The Cosmic ray
  connection}'', {\em Rept.Prog.Phys.} {\bfseries 65} (2002) 1025--1078,
 \href{http://xxx.lanl.gov/abs/astro-ph/0204527}{ arXiv:astro-ph/0204527}.

\bibitem[Lipari(2006)]{Lipari:2006uw}
P.~Lipari, ``{Perspectives of High Energy Neutrino Astronomy}'', {\em
  Nucl.Instrum.Meth.} {\bfseries A567} (2006) 405--417,
 \href{http://xxx.lanl.gov/abs/astro-ph/0605535}{ arXiv:astro-ph/0605535}.

\bibitem[Becker(2008)]{Becker:2007sv}
J.~K. Becker, ``{High-energy neutrinos in the context of multimessenger
  physics}'', {\em Phys.Rept.} {\bfseries 458} (2008) 173--246,
 \href{http://xxx.lanl.gov/abs/0710.1557}{ arXiv:0710.1557}.

\bibitem[Anchordoqui and Montaruli(2010)]{Anchordoqui:2009nf}
L.~A. Anchordoqui and T.~Montaruli, ``{In Search for Extraterrestrial High
  Energy Neutrinos}'', {\em Ann.Rev.Nucl.Part.Sci.} {\bfseries 60} (2010)
  129--162,
 \href{http://xxx.lanl.gov/abs/0912.1035}{ arXiv:0912.1035}.

\bibitem[Abbasi et~al.(2011)]{IceCube:2011ab}
{\bfseries IceCube} Collaboration, R.~Abbasi {\em et~al.}, ``{IceCube -
  Astrophysics and Astroparticle Physics at the South Pole}'',
 \href{http://xxx.lanl.gov/abs/1111.5188}{ arXiv:1111.5188}.

\bibitem[Cholis and Hooper(2013)]{Cholis:2012kq}
I.~Cholis and D.~Hooper, ``{On The Origin of IceCube's PeV Neutrinos}'', {\em
  JCAP} {\bfseries 06} (2013) 030,
 \href{http://xxx.lanl.gov/abs/1211.1974}{ arXiv:1211.1974}.

\bibitem[Liu and Wang(2013)]{Liu:2012pf}
R.-Y. Liu and X.-Y. Wang, ``{Diffuse PeV neutrinos from gamma-ray bursts}'',
  {\em Astrophys.J.} {\bfseries 766} (2013) 73,
 \href{http://xxx.lanl.gov/abs/1212.1260}{ arXiv:1212.1260}.

\bibitem[Kistler et~al.(2013)Kistler, Stanev, and Yuksel]{Kistler:2013my}
M.~D. Kistler, T.~Stanev, and H.~Yuksel, ``{Cosmic PeV Neutrinos and the
  Sources of Ultrahigh Energy Protons}'',
 \href{http://xxx.lanl.gov/abs/1301.1703}{ arXiv:1301.1703}.

\bibitem[He et~al.(2013)He, Wang, Fan, Liu, and Wei]{He:2013cqa}
H.-N. He, T.~Wang, Y.-Z. Fan, S.-M. Liu, and D.-M. Wei, ``{Diffuse PeV neutrino
  emission from Ultra-Luminous Infrared Galaxies}'',
 \href{http://xxx.lanl.gov/abs/1303.1253}{ arXiv:1303.1253}.

\bibitem[Kalashev et~al.(2013)Kalashev, Kusenko, and Essey]{Kalashev:2013vba}
O.~E. Kalashev, A.~Kusenko, and W.~Essey, ``{PeV neutrinos from intergalactic
  interactions of cosmic rays emitted by active galactic nuclei}'',
 \href{http://xxx.lanl.gov/abs/1303.0300}{ arXiv:1303.0300}.

\bibitem[Gupta(2013)]{Gupta:2013xfa}
N.~Gupta, ``{Galactic PeV Neutrinos}'',
 \href{http://xxx.lanl.gov/abs/1305.4123}{ arXiv:1305.4123}.

\bibitem[Fox et~al.(2013)Fox, Kashiyama, and Meszaros]{Fox:2013oza}
D.~Fox, K.~Kashiyama, and P.~Meszaros, ``{Sub-PeV Neutrinos from TeV
  Unidentified Sources in the Galaxy}'',
 \href{http://xxx.lanl.gov/abs/1305.6606}{ arXiv:1305.6606}.

\bibitem[Stecker(2013)]{Stecker:2013fxa}
F.~W. Stecker, ``{Ice Cube Observed PeV Neutrinos from AGN Cores}'',
 \href{http://xxx.lanl.gov/abs/1305.7404}{ arXiv:1305.7404}.

\bibitem[Anchordoqui et~al.(2013)Anchordoqui, Goldberg, Lynch, Olinto, Paul,
  et~al.]{Anchordoqui:2013lna}
L.~A. Anchordoqui, H.~Goldberg, M.~H. Lynch, A.~V. Olinto, T.~C. Paul, {\em
  et~al.}, ``{Pinning down the cosmic ray source mechanism with new IceCube
  data}'',
 \href{http://xxx.lanl.gov/abs/1306.5021}{ arXiv:1306.5021}.

\bibitem[Murase et~al.(2013)Murase, Ahlers, and Lacki]{Murase:2013rfa}
K.~Murase, M.~Ahlers, and B.~C. Lacki, ``{On the Hadronuclear Origin of PeV
  Neutrinos Observed with IceCube}'',
 \href{http://xxx.lanl.gov/abs/1306.3417}{ arXiv:1306.3417}.

\bibitem[Gao et~al.(2013)Gao, Zhang, Wu, and Dai]{Gao:2013rxa}
H.~Gao, B.~Zhang, X.-F. Wu, and Z.-G. Dai, ``{Possible High-Energy Neutrino and
  Photon Signals from Gravitational Wave Bursts due to Double Neutron Star
  Mergers}'',
 \href{http://xxx.lanl.gov/abs/1306.3006}{ arXiv:1306.3006}.

\bibitem[Murase and Ioka(2013)]{Murase:2013ffa}
K.~Murase and K.~Ioka, ``{TeV-PeV Neutrinos from Low-Power Gamma-Ray Burst Jets
  inside Stars}'',
 \href{http://xxx.lanl.gov/abs/1306.2274}{ arXiv:1306.2274}.

\bibitem[Roulet et~al.(2013)Roulet, Sigl, van Vliet, and
  Mollerach]{Roulet:2012rv}
E.~Roulet, G.~Sigl, A.~van Vliet, and S.~Mollerach, ``{PeV neutrinos from the
  propagation of ultra-high energy cosmic rays}'', {\em JCAP} {\bfseries 1301}
  (2013) 028,
 \href{http://xxx.lanl.gov/abs/1209.4033}{ arXiv:1209.4033}.

\bibitem[Fargion(2012{\natexlab{a}})]{Fargion:2012mw}
D.~Fargion, ``{Radioactive UHECR Astronomy: Correlating gamma anisotropy and
  neutrino PeV events}'',
 \href{http://xxx.lanl.gov/abs/1209.6090}{ arXiv:1209.6090}.

\bibitem[Fargion(2012{\natexlab{b}})]{Fargion:2012zx}
D.~Fargion, ``{TeV gamma-UHECR anisotropy by decaying nuclei in flight: first
  neutrino traces?}'',
 \href{http://xxx.lanl.gov/abs/1207.0254}{ arXiv:1207.0254}.

\bibitem[Fargion et~al.(2012)Fargion, D'Armiento, and Paggi]{Fargion:2012zc}
D.~Fargion, D.~D'Armiento, and P.~Paggi, ``{UHECR bending, clustering and
  decaying feeding gamma anisotropy}'',
 \href{http://xxx.lanl.gov/abs/1208.2471}{ arXiv:1208.2471}.

\bibitem[Pakvasa et~al.(2012)Pakvasa, Joshipura, and Mohanty]{Pakvasa:2012db}
S.~Pakvasa, A.~Joshipura, and S.~Mohanty, ``{Explanation for the low flux of
  high energy astrophysical muon-neutrinos}'',
 \href{http://xxx.lanl.gov/abs/1209.5630}{ arXiv:1209.5630}.

\bibitem[Baerwald et~al.(2012)Baerwald, Bustamante, and
  Winter]{Baerwald:2012kc}
P.~Baerwald, M.~Bustamante, and W.~Winter, ``{Neutrino Decays over Cosmological
  Distances and the Implications for Neutrino Telescopes}'', {\em JCAP}
  {\bfseries 1210} (2012) 020,
 \href{http://xxx.lanl.gov/abs/1208.4600}{ arXiv:1208.4600}.

\bibitem[Bhattacharya et~al.(2012)Bhattacharya, Gandhi, Rodejohann, and
  Watanabe]{Bhattacharya:2012fh}
A.~Bhattacharya, R.~Gandhi, W.~Rodejohann, and A.~Watanabe, ``{On the
  interpretation of IceCube cascade events in terms of the Glashow
  resonance}'',
 \href{http://xxx.lanl.gov/abs/1209.2422}{ arXiv:1209.2422}.

\bibitem[Feldstein et~al.(2013)Feldstein, Kusenko, Matsumoto, and
  Yanagida]{Feldstein:2013kka}
B.~Feldstein, A.~Kusenko, S.~Matsumoto, and T.~T. Yanagida, ``{Neutrinos at
  IceCube from Heavy Decaying Dark Matter}'',
 \href{http://xxx.lanl.gov/abs/1303.7320}{ arXiv:1303.7320}.

\bibitem[Barger and Keung(2013)]{Barger:2013pla}
V.~Barger and W.-Y. Keung, ``{Superheavy Particle Origin of IceCube PeV
  Neutrino Events}'',
 \href{http://xxx.lanl.gov/abs/1305.6907}{ arXiv:1305.6907}.

\bibitem[Borriello et~al.(2013)Borriello, Chakraborty, Mirizzi, and
  Serpico]{Borriello:2013ala}
E.~Borriello, S.~Chakraborty, A.~Mirizzi, and P.~D. Serpico, ``{A stringent
  constraint on neutrino Lorentz invariance violation from the two IceCube PeV
  neutrinos}'',
 \href{http://xxx.lanl.gov/abs/1303.5843}{ arXiv:1303.5843}.

\bibitem[Vissani et~al.(2013)Vissani, Pagliaroli, and
  Villante]{Vissani:2013iga}
F.~Vissani, G.~Pagliaroli, and F.~L. Villante, ``{The fraction of muon tracks
  in cosmic neutrinos}'',
 \href{http://xxx.lanl.gov/abs/1306.0211}{ arXiv:1306.0211}.

\bibitem[Stecker(2013)]{Stecker:2013jfa}
F.~W. Stecker, ``{Constraining Neutrino Velocities and Lorentz Invariance
  Violation in the Neutrino Sector using the IceCube PeV Neutrino Events}'',
 \href{http://xxx.lanl.gov/abs/1306.6095}{ arXiv:1306.6095}.

\bibitem[Laha(February 2013)]{LahaAspen:2013}
R.~Laha, ``{PeV cascades in IceCube: atmospheric prompt or astrophysical
  neutrinos?}'', {\em Aspen Winter Workshop--New Directions in Neutrino Physics
  \url{http://indico.cern.ch/conferenceDisplay.py?confId=224351}}, February
  2013.

\bibitem[Laha(May 2013)]{LahaIPA:2013}
R.~Laha, ``{PeV cascades in IceCube: the way ahead}'', {\em IceCube Particle
  Astrophysics (IPA) Symposium
  \url{http://events.icecube.wisc.edu/conferenceDisplay.py?confId=46}}, May
  2013.

\bibitem[Beacom and Candia(2004)]{Beacom:2004jb}
J.~F. Beacom and J.~Candia, ``{Shower power: Isolating the prompt atmospheric
  neutrino flux using electron neutrinos}'', {\em JCAP} {\bfseries 0411} (2004)
  009,
 \href{http://xxx.lanl.gov/abs/hep-ph/0409046}{ arXiv:hep-ph/0409046}.

\bibitem[Whitehorn(2012)]{Whitehorn:tevpa2012}
N.~Whitehorn, ``{Observations of PeV neutrinos in IceCube}'', {\em TeVPA 2012
  Conference
  \url{https://grapes-3.tifr.res.in/indico/internalPage.py?pageId=0&confId=0}},
  2012.

\bibitem[Neilson(2013)]{Neilson:Pheno2013}
N.~K. Neilson, ``{News from IceCube}'', {\em Pheno 2013 Symposium
  \url{http://indico.cern.ch/conferenceDisplay.py?confId=221653}}, 2013.

\bibitem[Abbasi et~al.(2010)]{Abbasi:2010ak}
{\bfseries IceCube} Collaboration, R.~Abbasi {\em et~al.}, ``{The first search
  for extremely-high energy cosmogenic neutrinos with the IceCube Neutrino
  Observatory}'', {\em Phys.Rev.} {\bfseries D82} (2010) 072003,
 \href{http://xxx.lanl.gov/abs/1009.1442}{ arXiv:1009.1442}.

\bibitem[Abbasi et~al.(2011)]{Abbasi:2011ji}
{\bfseries IceCube} Collaboration, R.~Abbasi {\em et~al.}, ``{Constraints on
  the Extremely-high Energy Cosmic Neutrino Flux with the IceCube 2008-2009
  Data}'', {\em Phys.Rev.} {\bfseries D83} (2011) 092003,
 \href{http://xxx.lanl.gov/abs/1103.4250}{ arXiv:1103.4250}.

\bibitem[Schonert et~al.(2009)Schonert, Gaisser, Resconi, and
  Schulz]{Schonert:2008is}
S.~Schonert, T.~K. Gaisser, E.~Resconi, and O.~Schulz, ``{Vetoing atmospheric
  neutrinos in a high energy neutrino telescope}'', {\em Phys.Rev.} {\bfseries
  D79} (2009) 043009,
 \href{http://xxx.lanl.gov/abs/0812.4308}{ arXiv:0812.4308}.

\bibitem[Honda et~al.(2007)Honda, Kajita, Kasahara, Midorikawa, and
  Sanuki]{Honda:2006qj}
M.~Honda, T.~Kajita, K.~Kasahara, S.~Midorikawa, and T.~Sanuki, ``{Calculation
  of atmospheric neutrino flux using the interaction model calibrated with
  atmospheric muon data}'', {\em Phys.Rev.} {\bfseries D75} (2007) 043006,
 \href{http://xxx.lanl.gov/abs/astro-ph/0611418}{ arXiv:astro-ph/0611418}.

\bibitem[Abbasi et~al.(2011)]{Abbasi:2011jx}
{\bfseries IceCube} Collaboration, R.~Abbasi {\em et~al.}, ``{A Search for a
  Diffuse Flux of Astrophysical Muon Neutrinos with the IceCube 40-String
  Detector}'', {\em Phys.Rev.} {\bfseries D84} (2011) 082001,
 \href{http://xxx.lanl.gov/abs/1104.5187}{ arXiv:1104.5187}.

\bibitem[Enberg et~al.(2008)Enberg, Reno, and Sarcevic]{Enberg:2008te}
R.~Enberg, M.~H. Reno, and I.~Sarcevic, ``{Prompt neutrino fluxes from
  atmospheric charm}'', {\em Phys.Rev.} {\bfseries D78} (2008) 043005,
 \href{http://xxx.lanl.gov/abs/0806.0418}{ arXiv:0806.0418}.

\bibitem[Takami et~al.(2009)Takami, Murase, Nagataki, and Sato]{Takami:2007pp}
H.~Takami, K.~Murase, S.~Nagataki, and K.~Sato, ``{Cosmogenic neutrinos as a
  probe of the transition from Galactic to extragalactic cosmic rays}'', {\em
  Astropart.Phys.} {\bfseries 31} (2009) 201--211,
 \href{http://xxx.lanl.gov/abs/0704.0979}{ arXiv:0704.0979}.

\bibitem[Ahlers et~al.(2010)Ahlers, Anchordoqui, Gonzalez-Garcia, Halzen, and
  Sarkar]{Ahlers:2010fw}
M.~Ahlers, L.~Anchordoqui, M.~Gonzalez-Garcia, F.~Halzen, and S.~Sarkar, ``{GZK
  Neutrinos after the Fermi-LAT Diffuse Photon Flux Measurement}'', {\em
  Astropart.Phys.} {\bfseries 34} (2010) 106--115,
 \href{http://xxx.lanl.gov/abs/1005.2620}{ arXiv:1005.2620}.

\bibitem[Aartsen et~al.(2012)]{Aartsen:2012uu}
{\bfseries IceCube} Collaboration, M.~Aartsen {\em et~al.}, ``{Measurement of
  the Atmospheric $\nu_e$ flux in IceCube}'',
 \href{http://xxx.lanl.gov/abs/1212.4760}{ arXiv:1212.4760}.

\bibitem[Abbasi et~al.(2011)]{Abbasi:2010ie}
{\bfseries IceCube} Collaboration, R.~Abbasi {\em et~al.}, ``{Measurement of
  the atmospheric neutrino energy spectrum from 100 GeV to 400 TeV with
  IceCube}'', {\em Phys.Rev.} {\bfseries D83} (2011) 012001,
 \href{http://xxx.lanl.gov/abs/1010.3980}{ arXiv:1010.3980}.

\bibitem[Gandhi et~al.(1998)Gandhi, Quigg, Reno, and Sarcevic]{Gandhi:1998ri}
R.~Gandhi, C.~Quigg, M.~H. Reno, and I.~Sarcevic, ``{Neutrino interactions at
  ultrahigh-energies}'', {\em Phys.Rev.} {\bfseries D58} (1998) 093009,
 \href{http://xxx.lanl.gov/abs/hep-ph/9807264}{ arXiv:hep-ph/9807264}.

\bibitem[Connolly et~al.(2011)Connolly, Thorne, and Waters]{Connolly:2011vc}
A.~Connolly, R.~S. Thorne, and D.~Waters, ``{Calculation of High Energy
  Neutrino-Nucleon Cross Sections and Uncertainties Using the MSTW Parton
  Distribution Functions and Implications for Future Experiments}'', {\em
  Phys.Rev.} {\bfseries D83} (2011) 113009,
 \href{http://xxx.lanl.gov/abs/1102.0691}{ arXiv:1102.0691}.

\bibitem[Cooper-Sarkar et~al.(2011)Cooper-Sarkar, Mertsch, and
  Sarkar]{CooperSarkar:2011pa}
A.~Cooper-Sarkar, P.~Mertsch, and S.~Sarkar, ``{The high energy neutrino
  cross-section in the Standard Model and its uncertainty}'', {\em JHEP}
  {\bfseries 1108} (2011) 042,
 \href{http://xxx.lanl.gov/abs/1106.3723}{ arXiv:1106.3723}.

\bibitem[Block et~al.(2013)Block, Durand, Ha, and McKay]{Block:2013nia}
M.~M. Block, L.~Durand, P.~Ha, and D.~W. McKay, ``{Implications of a Froissart
  bound saturation of $\gamma^*$-$p$ deep inelastic scattering. Part II.
  Ultra-high energy neutrino interactions}'',
 \href{http://xxx.lanl.gov/abs/1302.6127}{ arXiv:1302.6127}.

\bibitem[Gandhi et~al.(1996)Gandhi, Quigg, Reno, and Sarcevic]{Gandhi:1995tf}
R.~Gandhi, C.~Quigg, M.~H. Reno, and I.~Sarcevic, ``{Ultrahigh-energy neutrino
  interactions}'', {\em Astropart.Phys.} {\bfseries 5} (1996) 81--110,
 \href{http://xxx.lanl.gov/abs/hep-ph/9512364}{ arXiv:hep-ph/9512364}.

\bibitem[Glashow(1960)]{Glashow:1960zz}
S.~L. Glashow, ``{Resonant Scattering of Antineutrinos}'', {\em Phys.Rev.}
  {\bfseries 118} (1960)
316--317.

\bibitem[Barger et~al.(2012)Barger, Learned, and Pakvasa]{Barger:2012mz}
V.~Barger, J.~Learned, and S.~Pakvasa, ``{IceCube PeV Cascade Events Initiated
  by Electron-Antineutrinos at Glashow Resonance}'',
 \href{http://xxx.lanl.gov/abs/1207.4571}{ arXiv:1207.4571}.

\bibitem[Learned and Pakvasa(1995)]{Learned:1994wg}
J.~G. Learned and S.~Pakvasa, ``{Detecting tau-neutrino oscillations at PeV
  energies}'', {\em Astropart.Phys.} {\bfseries 3} (1995) 267--274,
 \href{http://xxx.lanl.gov/abs/hep-ph/9405296}{ arXiv:hep-ph/9405296}.

\bibitem[Beacom et~al.(2003)Beacom, Bell, Hooper, Pakvasa, and
  Weiler]{Beacom:2003nh}
J.~F. Beacom, N.~F. Bell, D.~Hooper, S.~Pakvasa, and T.~J. Weiler, ``{Measuring
  flavor ratios of high-energy astrophysical neutrinos}'', {\em Phys.Rev.}
  {\bfseries D68} (2003) 093005,
 \href{http://xxx.lanl.gov/abs/hep-ph/0307025}{ arXiv:hep-ph/0307025}.

\bibitem[Ritz and Seckel(1988)]{Ritz:1987mh}
S.~Ritz and D.~Seckel, ``{Detailed neutrino spectra from cold dark matter
  annihilations in the Sun}'', {\em Nucl.Phys.} {\bfseries B304} (1988)
877.

\bibitem[Halzen and Saltzberg(1998)]{Halzen:1998be}
F.~Halzen and D.~Saltzberg, ``{Tau-neutrino appearance with a 1000 megaparsec
  baseline}'', {\em Phys.Rev.Lett.} {\bfseries 81} (1998) 4305--4308,
 \href{http://xxx.lanl.gov/abs/hep-ph/9804354}{ arXiv:hep-ph/9804354}.

\bibitem[Greisen(1966)]{Greisen:1966jv}
K.~Greisen, ``{End to the cosmic ray spectrum?}'', {\em Phys.Rev.Lett.}
  {\bfseries 16} (1966)
748--750.

\bibitem[Zatsepin and Kuzmin(1966)]{Zatsepin:1966jv}
G.~Zatsepin and V.~Kuzmin, ``{Upper limit of the spectrum of cosmic rays}'',
  {\em JETP Lett.} {\bfseries 4} (1966)
78--80.

\bibitem[Berezinsky and Zatsepin(1970)]{Berezinsky:1970xj}
V.~Berezinsky and G.~Zatsepin, ``{Cosmic neutrinos of superhigh energy}'', {\em
  Yad.Fiz.} {\bfseries 11} (1970)
200--205.

\bibitem[Berezinsky and Zatsepin(1969)]{Beresinsky:1969qj}
V.~Berezinsky and G.~Zatsepin, ``{Cosmic rays at ultrahigh-energies
  (neutrino?)}'', {\em Phys.Lett.} {\bfseries B28} (1969)
423--424.

\bibitem[Yoshida and Teshima(1993)]{Yoshida:1993pt}
S.~Yoshida and M.~Teshima, ``{Energy spectrum of ultrahigh-energy cosmic rays
  with extragalactic origin}'', {\em Prog.Theor.Phys.} {\bfseries 89} (1993)
833--845.

\bibitem[Engel et~al.(2001)Engel, Seckel, and Stanev]{Engel:2001hd}
R.~Engel, D.~Seckel, and T.~Stanev, ``{Neutrinos from propagation of
  ultrahigh-energy protons}'', {\em Phys.Rev.} {\bfseries D64} (2001) 093010,
 \href{http://xxx.lanl.gov/abs/astro-ph/0101216}{ arXiv:astro-ph/0101216}.

\bibitem[Allard et~al.(2006)Allard, Ave, Busca, Malkan, Olinto,
  et~al.]{Allard:2006mv}
D.~Allard, M.~Ave, N.~Busca, M.~Malkan, A.~Olinto, {\em et~al.}, ``{Cosmogenic
  Neutrinos from the propagation of Ultrahigh Energy Nuclei}'', {\em JCAP}
  {\bfseries 0609} (2006) 005,
 \href{http://xxx.lanl.gov/abs/astro-ph/0605327}{ arXiv:astro-ph/0605327}.

\bibitem[Anchordoqui et~al.(2007)Anchordoqui, Goldberg, Hooper, Sarkar, and
  Taylor]{Anchordoqui:2007fi}
L.~A. Anchordoqui, H.~Goldberg, D.~Hooper, S.~Sarkar, and A.~M. Taylor,
  ``{Predictions for the Cosmogenic Neutrino Flux in Light of New Data from the
  Pierre Auger Observatory}'', {\em Phys.Rev.} {\bfseries D76} (2007) 123008,
 \href{http://xxx.lanl.gov/abs/0709.0734}{ arXiv:0709.0734}.

\bibitem[Kotera et~al.(2010)Kotera, Allard, and Olinto]{Kotera:2010yn}
K.~Kotera, D.~Allard, and A.~Olinto, ``{Cosmogenic Neutrinos: parameter space
  and detectabilty from PeV to ZeV}'', {\em JCAP} {\bfseries 1010} (2010) 013,
 \href{http://xxx.lanl.gov/abs/1009.1382}{ arXiv:1009.1382}.

\bibitem[Murase and Beacom(2010)]{Murase:2010gj}
K.~Murase and J.~F. Beacom, ``{Neutrino Background Flux from Sources of
  Ultrahigh-Energy Cosmic-Ray Nuclei}'', {\em Phys.Rev.} {\bfseries D81} (2010)
  123001,
 \href{http://xxx.lanl.gov/abs/1003.4959}{ arXiv:1003.4959}.

\bibitem[Gaisser(1990)]{Gaisser:1990vg}
T.~Gaisser, ``{Cosmic rays and particle physics}'',
1990.

\bibitem[Volkova(1980)]{Volkova:1980sw}
L.~Volkova, ``{Energy Spectra and Angular Distributions of Atmospheric
  Neutrinos}'', {\em Sov.J.Nucl.Phys.} {\bfseries 31} (1980)
784--790.

\bibitem[Gondolo et~al.(1996)Gondolo, Ingelman, and Thunman]{Gondolo:1995fq}
P.~Gondolo, G.~Ingelman, and M.~Thunman, ``{Charm production and high-energy
  atmospheric muon and neutrino fluxes}'', {\em Astropart.Phys.} {\bfseries 5}
  (1996) 309--332,
 \href{http://xxx.lanl.gov/abs/hep-ph/9505417}{ arXiv:hep-ph/9505417}.

\bibitem[Battistoni et~al.(1996)Battistoni, Bloise, Forti, Greco, Ranft,
  et~al.]{Battistoni:1995yv}
G.~Battistoni, C.~Bloise, C.~Forti, M.~Greco, J.~Ranft, {\em et~al.},
  ``{Calculation of the TeV prompt muon component in very high-energy cosmic
  ray showers}'', {\em Astropart.Phys.} {\bfseries 4} (1996)
351--364.

\bibitem[Pasquali et~al.(1999)Pasquali, Reno, and Sarcevic]{Pasquali:1998ji}
L.~Pasquali, M.~Reno, and I.~Sarcevic, ``{Lepton fluxes from atmospheric
  charm}'', {\em Phys.Rev.} {\bfseries D59} (1999) 034020,
 \href{http://xxx.lanl.gov/abs/hep-ph/9806428}{ arXiv:hep-ph/9806428}.

\bibitem[Gelmini et~al.(2000{\natexlab{a}})Gelmini, Gondolo, and
  Varieschi]{Gelmini:1999ve}
G.~Gelmini, P.~Gondolo, and G.~Varieschi, ``{Prompt atmospheric neutrinos and
  muons: NLO versus LO QCD predictions}'', {\em Phys.Rev.} {\bfseries D61}
  (2000){\natexlab{a}} 036005,
 \href{http://xxx.lanl.gov/abs/hep-ph/9904457}{ arXiv:hep-ph/9904457}.

\bibitem[Gelmini et~al.(2000{\natexlab{b}})Gelmini, Gondolo, and
  Varieschi]{Gelmini:1999xq}
G.~Gelmini, P.~Gondolo, and G.~Varieschi, ``{Prompt atmospheric neutrinos and
  muons: Dependence on the gluon distribution function}'', {\em Phys.Rev.}
  {\bfseries D61} (2000){\natexlab{b}} 056011,
 \href{http://xxx.lanl.gov/abs/hep-ph/9905377}{ arXiv:hep-ph/9905377}.

\bibitem[Martin et~al.(2003)Martin, Ryskin, and Stasto]{Martin:2003us}
A.~Martin, M.~Ryskin, and A.~Stasto, ``{Prompt neutrinos from atmospheric $c
  \bar{c}$ and $b \bar{b}$ production and the gluon at very small x}'', {\em
  Acta Phys.Polon.} {\bfseries B34} (2003) 3273--3304,
 \href{http://xxx.lanl.gov/abs/hep-ph/0302140}{ arXiv:hep-ph/0302140}.

\bibitem[Candia and Roulet(2003)]{Candia:2003ay}
J.~Candia and E.~Roulet, ``{Rigidity dependent knee and cosmic ray induced high
  energy neutrino fluxes}'', {\em JCAP} {\bfseries 0309} (2003) 005,
 \href{http://xxx.lanl.gov/abs/astro-ph/0306632}{ arXiv:astro-ph/0306632}.

\bibitem[Berghaus et~al.(2008)Berghaus, Montaruli, and Ranft]{Berghaus:2007hp}
P.~Berghaus, T.~Montaruli, and J.~Ranft, ``{Charm Production in DPMJET}'', {\em
  JCAP} {\bfseries 0806} (2008) 003,
 \href{http://xxx.lanl.gov/abs/0712.3089}{ arXiv:0712.3089}.

\bibitem[Sinegovskaya et~al.(2013)Sinegovskaya, Ogorodnikova, and
  Sinegovsky]{Sinegovskaya:2013iaa}
T.~Sinegovskaya, E.~Ogorodnikova, and S.~Sinegovsky, ``{High-energy fluxes of
  atmospheric neutrinos}'',
 \href{http://xxx.lanl.gov/abs/1306.5907}{ arXiv:1306.5907}.

\bibitem[Itow(2011)]{Itow:2011zz}
{\bfseries LHCf} Collaboration, Y.~Itow, ``{The first year results of the LHCf
  experiment to verify cosmic ray interaction models at LHC energy}'', {\em
  Prog.Theor.Phys.Suppl.} {\bfseries 187} (2011)
273--280.

\bibitem[Albacete et~al.(2012)Albacete, Milhano, Quiroga-Arias, and
  Rojo]{Albacete:2012rx}
J.~Albacete, J.~Milhano, P.~Quiroga-Arias, and J.~Rojo, ``{Linear vs Non-Linear
  QCD Evolution: From HERA Data to LHC Phenomenology}'', {\em Eur.Phys.J.}
  {\bfseries C72} (2012) 2131,
 \href{http://xxx.lanl.gov/abs/1203.1043}{ arXiv:1203.1043}.

\bibitem[Roland(2012)]{Roland:2012dc}
{\bfseries CMS} Collaboration, B.~Roland, ``{Recent CMS Results on Forward and
  Small-x QCD Physics}'',
 \href{http://xxx.lanl.gov/abs/1209.5632}{ arXiv:1209.5632}.

\bibitem[Naumov et~al.(1998)Naumov, Sinegovskaya, and
  Sinegovsky]{Naumov:1998vi}
V.~A. Naumov, T.~Sinegovskaya, and S.~Sinegovsky, ``{The K(lepton
  3)form-factors and atmospheric neutrino flavor ratio at high-energies}'',
  {\em Nuovo Cim.} {\bfseries A111} (1998) 129--148,
 \href{http://xxx.lanl.gov/abs/hep-ph/9802410}{ arXiv:hep-ph/9802410}.

\bibitem[Bugaev et~al.(1998)Bugaev, Misaki, Naumov, Sinegovskaya, Sinegovsky,
  et~al.]{Bugaev:1998bi}
E.~Bugaev, A.~Misaki, V.~A. Naumov, T.~Sinegovskaya, S.~Sinegovsky, {\em
  et~al.}, ``{Atmospheric muon flux at sea level, underground and
  underwater}'', {\em Phys.Rev.} {\bfseries D58} (1998) 054001,
 \href{http://xxx.lanl.gov/abs/hep-ph/9803488}{ arXiv:hep-ph/9803488}.

\bibitem[Zas et~al.(1993)Zas, Halzen, and Vazquez]{Zas:1992ci}
E.~Zas, F.~Halzen, and R.~Vazquez, ``{High-energy neutrino astronomy:
  Horizontal air shower arrays versus underground detectors}'', {\em
  Astropart.Phys.} {\bfseries 1} (1993)
297--316.

\bibitem[Schukraft(2013)]{Schukraft:2013ya}
{\bfseries IceCube} Collaboration, A.~Schukraft, ``{A view of prompt
  atmospheric neutrinos with IceCube}'', {\em Nucl.Phys.Proc.Suppl.}, 2013
 \href{http://xxx.lanl.gov/abs/1302.0127}{ arXiv:1302.0127}.

\bibitem[Gaisser(2013)]{Gaisser:2013ira}
T.~K. Gaisser, ``{Atmospheric leptons, the search for a prompt component}'',
 \href{http://xxx.lanl.gov/abs/1303.1431}{ arXiv:1303.1431}.

\bibitem[Blumer et~al.(2009)Blumer, Engel, and Horandel]{Bluemer:2009zf}
J.~Blumer, R.~Engel, and J.~R. Horandel, ``{Cosmic Rays from the Knee to the
  Highest Energies}'', {\em Prog.Part.Nucl.Phys.} {\bfseries 63} (2009)
  293--338,
 \href{http://xxx.lanl.gov/abs/0904.0725}{ arXiv:0904.0725}.

\bibitem[Gelmini et~al.(2003)Gelmini, Gondolo, and Varieschi]{Gelmini:2002sw}
G.~Gelmini, P.~Gondolo, and G.~Varieschi, ``{Measuring the prompt atmospheric
  neutrino flux with down-going muons in neutrino telescopes.}'', {\em
  Phys.Rev.} {\bfseries D67} (2003) 017301,
 \href{http://xxx.lanl.gov/abs/hep-ph/0209111}{ arXiv:hep-ph/0209111}.

\bibitem[Gandhi and Panda(2006)]{Gandhi:2005at}
R.~Gandhi and S.~Panda, ``{Probing the cosmic ray 'Knee' and very high energy
  prompt muon and neutrino fluxes via underground muons}'', {\em JCAP}
  {\bfseries 0607} (2006) 011,
 \href{http://xxx.lanl.gov/abs/hep-ph/0512179}{ arXiv:hep-ph/0512179}.

\bibitem[Desiati and Gaisser(2010)]{Desiati:2010wt}
P.~Desiati and T.~K. Gaisser, ``{Seasonal variation of atmospheric leptons as a
  probe of charm}'', {\em Phys.Rev.Lett.} {\bfseries 105} (2010) 121102,
 \href{http://xxx.lanl.gov/abs/1008.2211}{ arXiv:1008.2211}.

\bibitem[Mannheim(1995)]{Mannheim:1995mm}
K.~Mannheim, ``{High-energy neutrinos from extragalactic jets}'', {\em
  Astropart.Phys.} {\bfseries 3} (1995)
295--302.

\bibitem[Halzen and Zas(1997)]{Halzen:1997hw}
F.~Halzen and E.~Zas, ``{Neutrino fluxes from active galaxies: A Model
  independent estimate}'', {\em Astrophys.J.} {\bfseries 488} (1997) 669--674,
 \href{http://xxx.lanl.gov/abs/astro-ph/9702193}{ arXiv:astro-ph/9702193}.

\bibitem[Atoyan and Dermer(2001)]{Atoyan:2001ey}
A.~Atoyan and C.~D. Dermer, ``{High-energy neutrinos from photomeson processes
  in blazars}'', {\em Phys.Rev.Lett.} {\bfseries 87} (2001) 221102,
 \href{http://xxx.lanl.gov/abs/astro-ph/0108053}{ arXiv:astro-ph/0108053}.

\bibitem[Muecke et~al.(2003)Muecke, Protheroe, Engel, Rachen, and
  Stanev]{Muecke:2002bi}
A.~Muecke, R.~Protheroe, R.~Engel, J.~Rachen, and T.~Stanev, ``{BL Lac Objects
  in the synchrotron proton blazar model}'', {\em Astropart.Phys.} {\bfseries
  18} (2003) 593--613,
 \href{http://xxx.lanl.gov/abs/astro-ph/0206164}{ arXiv:astro-ph/0206164}.

\bibitem[Anchordoqui et~al.(2008)Anchordoqui, Hooper, Sarkar, and
  Taylor]{Anchordoqui:2007tn}
L.~A. Anchordoqui, D.~Hooper, S.~Sarkar, and A.~M. Taylor, ``{High-energy
  neutrinos from astrophysical accelerators of cosmic ray nuclei}'', {\em
  Astropart.Phys.} {\bfseries 29} (2008) 1--13,
 \href{http://xxx.lanl.gov/abs/astro-ph/0703001}{ arXiv:astro-ph/0703001}.

\bibitem[Stecker et~al.(1991)Stecker, Done, Salamon, and
  Sommers]{Stecker:1991vm}
F.~Stecker, C.~Done, M.~Salamon, and P.~Sommers, ``{High-energy neutrinos from
  active galactic nuclei}'', {\em Phys.Rev.Lett.} {\bfseries 66} (1991)
2697--2700.

\bibitem[Alvarez-Muniz and Meszaros(2004)]{AlvarezMuniz:2004uz}
J.~Alvarez-Muniz and P.~Meszaros, ``{High energy neutrinos from radio-quiet
  AGNs}'', {\em Phys.Rev.} {\bfseries D70} (2004) 123001,
 \href{http://xxx.lanl.gov/abs/astro-ph/0409034}{ arXiv:astro-ph/0409034}.

\bibitem[Waxman and Bahcall(1997)]{Waxman:1997ti}
E.~Waxman and J.~N. Bahcall, ``{High-energy neutrinos from cosmological
  gamma-ray burst fireballs}'', {\em Phys.Rev.Lett.} {\bfseries 78} (1997)
  2292--2295,
 \href{http://xxx.lanl.gov/abs/astro-ph/9701231}{ arXiv:astro-ph/9701231}.

\bibitem[Dermer and Atoyan(2003)]{Dermer:2003zv}
C.~D. Dermer and A.~Atoyan, ``{High energy neutrinos from gamma-ray bursts}'',
  {\em Phys.Rev.Lett.} {\bfseries 91} (2003) 071102,
 \href{http://xxx.lanl.gov/abs/astro-ph/0301030}{ arXiv:astro-ph/0301030}.

\bibitem[Murase and Nagataki(2006)]{Murase:2005hy}
K.~Murase and S.~Nagataki, ``{High energy neutrino emission and neutrino
  background from gamma-ray bursts in the internal shock model}'', {\em
  Phys.Rev.} {\bfseries D73} (2006) 063002,
 \href{http://xxx.lanl.gov/abs/astro-ph/0512275}{ arXiv:astro-ph/0512275}.

\bibitem[Waxman and Bahcall(2000)]{Waxman:1999ai}
E.~Waxman and J.~N. Bahcall, ``{Neutrino afterglow from gamma-ray bursts:
  Similar to 10**18-eV}'', {\em Astrophys.J.} {\bfseries 541} (2000) 707--711,
 \href{http://xxx.lanl.gov/abs/hep-ph/9909286}{ arXiv:hep-ph/9909286}.

\bibitem[Dermer(2002)]{Dermer:2000yd}
C.~D. Dermer, ``{Neutrino, neutron, and cosmic ray production in the external
  shock model of gamma-ray bursts}'', {\em Astrophys.J.} {\bfseries 574} (2002)
  65--87,
 \href{http://xxx.lanl.gov/abs/astro-ph/0005440}{ arXiv:astro-ph/0005440}.

\bibitem[Murase and Nagataki(2006)]{Murase:2006dr}
K.~Murase and S.~Nagataki, ``{High Energy Neutrino Flash from Far-UV/X-ray
  Flares of Gamma-Ray Bursts}'', {\em Phys.Rev.Lett.} {\bfseries 97} (2006)
  051101,
 \href{http://xxx.lanl.gov/abs/astro-ph/0604437}{ arXiv:astro-ph/0604437}.

\bibitem[Murase et~al.(2009)Murase, Meszaros, and Zhang]{Murase:2009pg}
K.~Murase, P.~Meszaros, and B.~Zhang, ``{Probing the birth of fast rotating
  magnetars through high-energy neutrinos}'', {\em Phys.Rev.} {\bfseries D79}
  (2009) 103001,
 \href{http://xxx.lanl.gov/abs/0904.2509}{ arXiv:0904.2509}.

\bibitem[Murase et~al.(2011)Murase, Thompson, Lacki, and Beacom]{Murase:2010cu}
K.~Murase, T.~A. Thompson, B.~C. Lacki, and J.~F. Beacom, ``{New Class of
  High-Energy Transients from Crashes of Supernova Ejecta with Massive
  Circumstellar Material Shells}'', {\em Phys.Rev.} {\bfseries D84} (2011)
  043003,
 \href{http://xxx.lanl.gov/abs/1012.2834}{ arXiv:1012.2834}.

\bibitem[Katz et~al.(2011)Katz, Sapir, and Waxman]{Katz:2011zx}
B.~Katz, N.~Sapir, and E.~Waxman, ``{X-rays, gamma-rays and neutrinos from
  collisoinless shocks in supernova wind breakouts}'',
 \href{http://xxx.lanl.gov/abs/1106.1898}{ arXiv:1106.1898}.

\bibitem[Loeb and Waxman(2006)]{Loeb:2006tw}
A.~Loeb and E.~Waxman, ``{The Cumulative background of high energy neutrinos
  from starburst galaxies}'', {\em JCAP} {\bfseries 0605} (2006) 003,
 \href{http://xxx.lanl.gov/abs/astro-ph/0601695}{ arXiv:astro-ph/0601695}.

\bibitem[Thompson et~al.(2006)Thompson, Quataert, Waxman, and
  Loeb]{Thompson:2006np}
T.~A. Thompson, E.~Quataert, E.~Waxman, and A.~Loeb, ``{Assessing The Starburst
  Contribution to the Gamma-Ray and Neutrino Backgrounds}'',
 \href{http://xxx.lanl.gov/abs/astro-ph/0608699}{ arXiv:astro-ph/0608699}.

\bibitem[Berezinsky et~al.(1996)Berezinsky, Blasi, and
  Ptuskin]{Berezinsky:1996wx}
V.~Berezinsky, P.~Blasi, and V.~Ptuskin, ``{Clusters of galaxies as a storage
  room for cosmic rays}'', {\em Astrophys J.} {\bfseries 487} (1996) 529,
 \href{http://xxx.lanl.gov/abs/astro-ph/9609048}{ arXiv:astro-ph/9609048}.

\bibitem[Murase et~al.(2008)Murase, Inoue, and Nagataki]{Murase:2008yt}
K.~Murase, S.~Inoue, and S.~Nagataki, ``{Cosmic Rays Above the Second Knee from
  Clusters of Galaxies and Associated High-Energy Neutrino Emission}'', {\em
  Astrophys.J.} {\bfseries 689} (2008) L105,
 \href{http://xxx.lanl.gov/abs/0805.0104}{ arXiv:0805.0104}.

\bibitem[Kotera et~al.(2009)Kotera, Allard, Murase, Aoi, Dubois,
  et~al.]{Kotera:2009ms}
K.~Kotera, D.~Allard, K.~Murase, J.~Aoi, Y.~Dubois, {\em et~al.},
  ``{Propagation of ultrahigh energy nuclei in clusters of galaxies: resulting
  composition and secondary emissions}'', {\em Astrophys.J.} {\bfseries 707}
  (2009) 370--386,
 \href{http://xxx.lanl.gov/abs/0907.2433}{ arXiv:0907.2433}.

\bibitem[Kashti and Waxman(2005)]{Kashti:2005qa}
T.~Kashti and E.~Waxman, ``{Flavoring astrophysical neutrinos: Flavor ratios
  depend on energy}'', {\em Phys.Rev.Lett.} {\bfseries 95} (2005) 181101,
 \href{http://xxx.lanl.gov/abs/astro-ph/0507599}{ arXiv:astro-ph/0507599}.

\bibitem[Abbasi et~al.(2012)]{Abbasi:2012cu}
{\bfseries IceCube} Collaboration, R.~Abbasi {\em et~al.}, ``{A Search for UHE
  Tau Neutrinos with IceCube}'', {\em Phys.Rev.} {\bfseries D86} (2012) 022005,
 \href{http://xxx.lanl.gov/abs/1202.4564}{ arXiv:1202.4564}.

\bibitem[Waxman and Bahcall(1999)]{Waxman:1998yy}
E.~Waxman and J.~N. Bahcall, ``{High-energy neutrinos from astrophysical
  sources: An Upper bound}'', {\em Phys.Rev.} {\bfseries D59} (1999) 023002,
 \href{http://xxx.lanl.gov/abs/hep-ph/9807282}{ arXiv:hep-ph/9807282}.

\bibitem[Murase and Beacom(2012)]{Murase:2012xs}
K.~Murase and J.~F. Beacom, ``{Constraining Very Heavy Dark Matter Using
  Diffuse Backgrounds of Neutrinos and Cascaded Gamma Rays}'', {\em JCAP}
  {\bfseries 1210} (2012) 043,
 \href{http://xxx.lanl.gov/abs/1206.2595}{ arXiv:1206.2595}.

\bibitem[Essey et~al.(2010)Essey, Kalashev, Kusenko, and Beacom]{Essey:2009ju}
W.~Essey, O.~E. Kalashev, A.~Kusenko, and J.~F. Beacom, ``{Secondary photons
  and neutrinos from cosmic rays produced by distant blazars}'', {\em
  Phys.Rev.Lett.} {\bfseries 104} (2010) 141102,
 \href{http://xxx.lanl.gov/abs/0912.3976}{ arXiv:0912.3976}.

\bibitem[Essey et~al.(2011)Essey, Kalashev, Kusenko, and Beacom]{Essey:2010er}
W.~Essey, O.~Kalashev, A.~Kusenko, and J.~F. Beacom, ``{Role of line-of-sight
  cosmic ray interactions in forming the spectra of distant blazars in TeV
  gamma rays and high-energy neutrinos}'', {\em Astrophys.J.} {\bfseries 731}
  (2011) 51,
 \href{http://xxx.lanl.gov/abs/1011.6340}{ arXiv:1011.6340}.

\bibitem[Feldman and Cousins(1998)]{Feldman:1997qc}
G.~J. Feldman and R.~D. Cousins, ``{A Unified approach to the classical
  statistical analysis of small signals}'', {\em Phys.Rev.} {\bfseries D57}
  (1998) 3873--3889,
 \href{http://xxx.lanl.gov/abs/physics/9711021}{ arXiv:physics/9711021}.

\bibitem[Kistler and Beacom(2006)]{Kistler:2006hp}
M.~D. Kistler and J.~F. Beacom, ``{Guaranteed and Prospective Galactic TeV
  Neutrino Sources}'', {\em Phys.Rev.} {\bfseries D74} (2006) 063007,
 \href{http://xxx.lanl.gov/abs/astro-ph/0607082}{ arXiv:astro-ph/0607082}.

\bibitem[Lipari and Stanev(1991)]{Lipari:1991ut}
P.~Lipari and T.~Stanev, ``{Propagation of multi - TeV muons}'', {\em
  Phys.Rev.} {\bfseries D44} (1991)
3543--3554.

\bibitem[Dutta et~al.(2001)Dutta, Reno, Sarcevic, and Seckel]{Dutta:2000hh}
S.~I. Dutta, M.~Reno, I.~Sarcevic, and D.~Seckel, ``{Propagation of muons and
  taus at high-energies}'', {\em Phys.Rev.} {\bfseries D63} (2001) 094020,
 \href{http://xxx.lanl.gov/abs/hep-ph/0012350}{ arXiv:hep-ph/0012350}.

\bibitem[Barwick et~al.(1992)Barwick, Halzen, Lowder, Miller, Morse,
  et~al.]{Barwick:1991ur}
S.~Barwick, F.~Halzen, D.~Lowder, T.~Miller, R.~Morse, {\em et~al.},
  ``{Neutrino astronomy on the 1-KM**2 scale}'', {\em J.Phys.} {\bfseries G18}
  (1992)
225--248.

\bibitem[Lowder et~al.(1991)Lowder, Miller, Price, Westphal, Barwick,
  et~al.]{Lowder:1991uy}
D.~Lowder, T.~Miller, P.~Price, A.~Westphal, S.~Barwick, {\em et~al.},
  ``{Observation of muons using the polar ice cap as a Cherenkov detector}'',
  {\em Nature} {\bfseries 353} (1991)
331--333.

\bibitem[Andres et~al.(2000)Andres, Askebjer, Barwick, Bay, Bergstrom,
  et~al.]{Andres:1999hm}
E.~Andres, P.~Askebjer, S.~Barwick, R.~Bay, L.~Bergstrom, {\em et~al.}, ``{The
  AMANDA neutrino telescope: Principle of operation and first results}'', {\em
  Astropart.Phys.} {\bfseries 13} (2000) 1--20,
 \href{http://xxx.lanl.gov/abs/astro-ph/9906203}{ arXiv:astro-ph/9906203}.

\bibitem[Ahrens et~al.(2004)]{Ahrens:2003ix}
{\bfseries IceCube} Collaboration, J.~Ahrens {\em et~al.}, ``{Sensitivity of
  the IceCube detector to astrophysical sources of high energy muon
  neutrinos}'', {\em Astropart.Phys.} {\bfseries 20} (2004) 507--532,
 \href{http://xxx.lanl.gov/abs/astro-ph/0305196}{ arXiv:astro-ph/0305196}.

\bibitem[Achterberg et~al.(2006)]{Achterberg:2006md}
{\bfseries IceCube} Collaboration, A.~Achterberg {\em et~al.}, ``{First Year
  Performance of The IceCube Neutrino Telescope}'', {\em Astropart.Phys.}
  {\bfseries 26} (2006) 155--173,
 \href{http://xxx.lanl.gov/abs/astro-ph/0604450}{ arXiv:astro-ph/0604450}.

\bibitem[Ahrens et~al.(2001)]{IceCube:2001}
{\bfseries IceCube} Collaboration, J.~Ahrens {\em et~al.}, ``{IceCube
  Preliminary Design Document}'', {\em
  \url{http://icecube.wisc.edu/icecube/static/reports/IceCubeDesignDoc.pdf}},
  2001.

\bibitem[Whitehorn et~al.(May 2013)Whitehorn, Kopper, and
  Neilson]{Whitehorn:IPA2013}
N.~Whitehorn, C.~Kopper, and N.~Neilson, ``{Results from IceCube}'', {\em
  IceCube Particle Astrophysics (IPA) Symposium
  \url{http://events.icecube.wisc.edu/conferenceDisplay.py?confId=46}}, May
  2013.

\bibitem[Rachen and Meszaros(1998)]{Rachen:1998fd}
J.~P. Rachen and P.~Meszaros, ``{Photohadronic neutrinos from transients in
  astrophysical sources}'', {\em Phys.Rev.} {\bfseries D58} (1998) 123005,
 \href{http://xxx.lanl.gov/abs/astro-ph/9802280}{ arXiv:astro-ph/9802280}.

\bibitem[Murase(2008)]{Murase:2008sp}
K.~Murase, ``{Prompt High-Energy Neutrinos from Gamma-Ray Bursts in the
  Photospheric and Synchrotron Self-Compton Scenarios}'', {\em Phys.Rev.}
  {\bfseries D78} (2008) 101302,
 \href{http://xxx.lanl.gov/abs/0807.0919}{ arXiv:0807.0919}.

\bibitem[Wang and Dai(2009)]{Wang:2008zm}
X.-Y. Wang and Z.-G. Dai, ``{Prompt TeV neutrinos from dissipative photospheres
  of gamma-ray bursts}'', {\em Astrophys.J.} {\bfseries 691} (2009) L67--L71,
 \href{http://xxx.lanl.gov/abs/0807.0290}{ arXiv:0807.0290}.

\bibitem[Abbasi et~al.(2012)]{Abbasi:2012zw}
{\bfseries IceCube} Collaboration, R.~Abbasi {\em et~al.}, ``{An absence of
  neutrinos associated with cosmic-ray acceleration in $\gamma$-ray bursts}'',
  {\em Nature} {\bfseries 484} (2012) 351--353,
 \href{http://xxx.lanl.gov/abs/1204.4219}{ arXiv:1204.4219}.

\bibitem[Hummer et~al.(2012)Hummer, Baerwald, and Winter]{Hummer:2011ms}
S.~Hummer, P.~Baerwald, and W.~Winter, ``{Neutrino Emission from Gamma-Ray
  Burst Fireballs, Revised}'', {\em Phys.Rev.Lett.} {\bfseries 108} (2012)
  231101,
 \href{http://xxx.lanl.gov/abs/1112.1076}{ arXiv:1112.1076}.

\bibitem[He et~al.(2012)He, Liu, Wang, Nagataki, Murase, et~al.]{He:2012tq}
H.-N. He, R.-Y. Liu, X.-Y. Wang, S.~Nagataki, K.~Murase, {\em et~al.},
  ``{Icecube non-detection of GRBs: Constraints on the fireball properties}'',
  {\em Astrophys.J.} {\bfseries 752} (2012) 29,
 \href{http://xxx.lanl.gov/abs/1204.0857}{ arXiv:1204.0857}.

\bibitem[Murase et~al.(2006)Murase, Ioka, Nagataki, and
  Nakamura]{Murase:2006mm}
K.~Murase, K.~Ioka, S.~Nagataki, and T.~Nakamura, ``{High Energy Neutrinos and
  Cosmic-Rays from Low-Luminosity Gamma-Ray Bursts?}'', {\em Astrophys.J.}
  {\bfseries 651} (2006) L5--L8,
 \href{http://xxx.lanl.gov/abs/astro-ph/0607104}{ arXiv:astro-ph/0607104}.

\bibitem[Gupta and Zhang(2007)]{Gupta:2006jm}
N.~Gupta and B.~Zhang, ``{Neutrino Spectra from Low and High Luminosity
  Populations of Gamma Ray Bursts}'', {\em Astropart.Phys.} {\bfseries 27}
  (2007) 386--391,
 \href{http://xxx.lanl.gov/abs/astro-ph/0606744}{ arXiv:astro-ph/0606744}.

\bibitem[Kelner et~al.(2006)Kelner, Aharonian, and Bugayov]{Kelner:2006tc}
S.~Kelner, F.~A. Aharonian, and V.~Bugayov, ``{Energy spectra of gamma-rays,
  electrons and neutrinos produced at proton-proton interactions in the very
  high energy regime}'', {\em Phys.Rev.} {\bfseries D74} (2006) 034018,
 \href{http://xxx.lanl.gov/abs/astro-ph/0606058}{ arXiv:astro-ph/0606058}.

\bibitem[Acero(2009)]{Acero:2009nb}
{\bfseries HESS} Collaboration, F.~Acero, ``{Detection of Gamma Rays From a
  Starburst Galaxy}'', {\em Science} {\bfseries 326} (2009) 1080,
 \href{http://xxx.lanl.gov/abs/0909.4651}{ arXiv:0909.4651}.

\end{thebibliography}\endgroup

\end{document}